\documentclass[fleqn,usenatbib]{mnras}

\usepackage{newtxtext,newtxmath}

\usepackage[T1]{fontenc}

\DeclareRobustCommand{\VAN}[3]{#2}
\let\VANthebibliography\thebibliography
\def\thebibliography{\DeclareRobustCommand{\VAN}[3]{##3}\VANthebibliography}

\usepackage{graphicx}	
\usepackage{amsmath}	
\usepackage{anyfontsize}
\usepackage[toc,page]{appendix}
\usepackage[hypcap=false]{caption}
\usepackage{ulem}
\usepackage{oplotsymbl}
\usepackage{MnSymbol}
\usepackage{xcolor}

\title[C$-$D$\rm_{oop}$ vibrational modes in PAHs]{Investigating C$-$D out-of-plane vibrational modes in PAHs as a tool to study interstellar deuterium-containing PAHs}

\author[Buragohain et al.]{
Mridusmita Buragohain$^{1}$%
\thanks{Contact e-mail: \href{ms.mridusmita@gmail.com (MB),}
{ms.mridusmita@gmail.com}, ms.mridusmita@uohyd.ac.in}, 
Takashi Onaka$^{2}$,
Amit Pathak$^{3}$, 
Akant Vats$^{4}$, 
Itsuki Sakon$^{5}$ \\
$^{1}$School of Physics, University of Hyderabad, Hyderabad 500 046, India\\
$^{2}$Department of Astronomy, Graduate School of Science, The University of Tokyo, Japan \\
$^{3}$Department of Physics, Banaras Hindu University, Varanasi 221 005, India \\
$^{4}$NASA AMES Research Center, Moffett Field, CA 94035, USA \\
$^{5}$Institute of Astronomy, Graduate School of Science, The University of Tokyo, Japan
}

\date{Accepted XXX. Received YYY; in original form ZZZ}

\pubyear{\the\year{}}

\begin{document}
\label{firstpage}
\pagerange{\pageref{firstpage}--\pageref{lastpage}}
\maketitle
\begin{abstract}
Previous as well as recent observations by \textit{ISO}, \textit{Spitzer}, \textit{AKARI}, \textit{SOFIA}, \textit{JWST} etc. have revealed various characteristics of mid-infrared emission bands between 3$-$20~$\rm \mu m$. Subsequently, several forms of organics including  Polycylic Aromatic Hydrocarbons (PAHs)/PAH-like molecules are proposed as carriers for these bands. Deuterated PAH (PAD) is one such substituted PAH, which is proposed as a potential candidate carrier for weak emission bands at 4.4 and 4.65~$\mu \rm m$, detected towards few astronomical targets and are characteristics of aromatic and aliphatic C$-$D stretching modes in a PAD molecule, respectively. 
However, the 4.4 $\mu$m band is not widely detected. In order to validate PADs as carriers for mid-infrared emission bands, an additional alternative {tool} is crucial. If PAHs are deuterated, they should also possess an inherent signature from the C$-$D out-of-plane (C$-$D$\rm_{oop}$) vibrations, which are at {the} longer wavelength side. 
In this report, features due to C$-$D$\rm_{oop}$ modes in PAHs {bearing} a single to multiple deuterium {atoms} 
are reported by performing quantum-chemical calculations. {This paper} reports that some of the C$-$D$\rm_{oop}$ vibrations appear at {the} $14-19$\,$\mu$m range.
Also, the strength of C$-$D$\rm_{oop}$ modes is not proportional to {the D/H ratio} in PAHs. {In addition, a moderate change in the spectra of deuterated PAHs is observed from that of the undeuterated counterparts, as deuteration would alternate the adjacency class of {the} C-H bonds and the symmetry of the molecule.} 
We discuss the efficiency and usefulness of these bands to constrain the form of PAHs emitting mid-infrared emission bands.
\end{abstract}



\section{Introduction}
Free flying gas phase Polycyclic Aromatic Hydrocarbon (PAH) molecules in the Interstellar Medium (ISM) have been so far the most attractive candidate carriers of the frequently detected broad mid-infrared emission bands, popularly known as `Aromatic Infrared Bands' (AIBs) \citep[]{Tielens08}. These bands are manifestations of vibrational modes in PAHs on absorption of background FUV photons \citep[]{Leger84, Allamandola85}.  While {regular} unsubstituted PAHs show roughly similar features, substitution of heteroatoms or even the slightest modification in the PAHs' geometry {(for example: five membered ring PAHs, dehydrogenated PAHs, superhyderogenated PAHs etc.)} can give rise to a set of new features or changes the intensity and position of the primary features, some of which {can} be even observationally detected. Deuterated PAH (PAD) \citep[]{Bausch97a, Hudgins04a} is one such substituted PAH, which is proposed as potential candidate carrier for weak bands at 4.4 and 4.65~$\mu \rm m$ towards {the} Orion Nebula and M17 \citep[]{Peeters04, Onaka14, Doney15} and are characteristics of  aromatic and aliphatic C$-$D stretching modes present in a PAD molecule, respectively
\citep[]{Mridu20, Yang2023}. {\citet[]{Onaka22} {also} reported 
excess emission at 4.4~$\mu \rm m$ towards galactic centre of a massing young stellar object (AFGL 2006).} JWST NIRSpec has recenlty detected a band at $\sim$~4.65~$\mu \rm m$, possibly from aliphatic deuterated PAHs, but it does not show any {clear} aromatic deuterated PAH signature at $\sim 4.4~\mu \rm m$ \citep[]{Boersma2023, peeters2024}.

The possible presence of deuterium-containing PAHs {has been discussed in relation to} the `problem of missing D', which suggests that some primordial missing D might be sequestered in PAHs \citep[]{Draine06, Allamandola21}.
Deuterium is one of the elements formed immediately after the Big Bang and {in course of} the chemical evolution of the Universe, nuclear fusion at the stellar interiors convert them into heavier elements. The lower value of [D/H]$\rm_{total(gas~plus~dust)}$ {at present  than} the {predicted} primordial value [D/H]$\rm_{prim}$ should be accounted for {solely} by chemical evolution model of our Galaxy. However, a wide range of [D/H]$\rm_{gas}$ has been observed along different lines of sight within our own Galaxy which cannot be explained by the chemical evolution model \citep[]{Linsky06}. In addition, the measured [D/H]$\rm_{gas}$ seems to be correlated with the interstellar depletion of heavy elements of Fe, Si, Ti, etc., suggesting D depletion onto dust grains. 
\citet[]{friedman2023} recently suggested that the spatial variability of D/H is not solely due to dust depletion. However, on account of the latest observations of deuterated features, some fraction of the missing D may still be attributable to dust depletion or depletion onto PAHs. Deuteration of PAHs may efficiently occur in low-temperature environments due to the difference in the zero-point energy \citep[]{Sandford2001}. However, the supposedly main diagnostic {signature} of deuterated PAHs {of} C$-$D stretching features at 4.4$-$4.65~$\mu \rm m$ may not be sufficiently excited in low-temperature environments as {they} may require high-energy photons to be seen in emission. This makes it difficult to probe the deuteration of PAHs from the study of 4.4$-$4.65~$\mu \rm m$ emission bands, even though D may be sequestered more in PAHs at low-temperatures
or only in large PAHs, both of which are difficult to be detected at 4~$\mu \rm m$. Thus,
the search for longer wavelength features arising from deuterium substitution may be important as suggested in \citet[]{Onaka22}.

Among the majorly observed AIBs, {the feature at 11.2~$\mu \rm m$} is one of the most distinctive and examined bands in the AIB spectrum \citep[]{Roche89, Roelfsema96, Sloan99, Hony01, Diedenhoven04, pasquini2024} and has been assigned to the C$-$H out-of -plane (C$-$H$\rm_{oop}$) bending mode 
of a neutral PAH molecule that contains a solo C$-$H group \citep[]{Cohen85, Witteborn89, Hudgins99}. A solo C$-$H group or non-adjacent C$-$H unit is one with no neighboring adjacent C-H units in the same ring. In a trend of increasing adjacency of C-H groups, PAHs are defined to have a duet (doubly-adjacent), trio (triply-adjacent), quartet (quadruply-adjacent) and quintet C$-$H (quintuply-adjacent) group \citep[]{Hudgins99}, each group giving rise to unique features. {The wavelength of C$-$H$\rm_{oop}$ vibration of a solo C$-$H group matches well with that of the observed emission feature at 11.2~$\mu \rm m$ indicating {the} abundance of PAHs with solo C$-$H group.} {A neighbouring feature at 11.0~$\mu \rm m$ is proposed to rise from solo C$-$H$\rm_{oop}$ modes in a PAH cation \citep[]{Hudgins99}.} If PAHs are deuterated or deuteronated, they should also have an inherent signature from the C$-$D out-of-plane (C$-$D$\rm_{oop}$) vibrations, which are at {the} longer wavelength side. Either these {could be} overlapping with other modes or are weak to be observationally detected. Studying these characteristic bands arising from {deuterium exchange} in PAHs can constrain {the abundance of} PAHs containing deuterium, if any. This work reports a Density Functional Theory (DFT) study on PAH molecules with a solo, duet, trio and quartet C$-$D groups to {investigate} the C$-$D$\rm_{oop}$ vibrational modes and to {show} how deuteration at different adjacency site{s} affects the spectra.  Even though D is an isotope of H, C$-$H and C$-$D are treated as separate groups for the purpose of the paper. A solo C$-$D unit is one with no neighboring adjacent C$-$D units in the same ring and with increasing adjacency of C$-$D units, these are named as duo, trio, quartet C$-$D units etc. The classification of these groups {is} made in similar ways as for C$-$H groups. {Note that oops modes are relatively easily excited even in low-temperature regions and/or arise from large PAHs.} The paper is arranged in the following manner; i) Theoretical Approach, ii) Results, where PAHs with each type (solo, duet, trio and quartet C$-$H/C$-$D groups) are discussed separately, iii) Discussion \& iv) Conclusion.
\section{Theoretical Approach}
Theoretical quantum chemical calculations, such as DFT has appeared to be {reliable} and time efficient for studying vibrational properties of PAHs, which is required to find correlation with the observed AIBs \citep[]{Hudgins01, Hudgins04a, Pathak05, Pathak06, Pathak07, Candian14, Mridu15}.  
For our purpose, we {perform}
DFT {calculations} combined with {the} B3LYP/6-311G** basis set to {optimize the} structure of the PAHs{,} which will further be used to obtain the harmonic frequencies and intensities of vibrational modes of the molecules at the same level of theory. In order to bring down the computed frequency which is usually overestimated compared to {the} experimental frequency, mode-dependent scaling factors 
are used \citep[]{Mridu15, Mridu16}. {The harmonic approximation with scaling factors of the oop modes provides sufficiently accurate wavelengths and intensities for the purpose of the present study \citep[]{mackie2022, lemmens2023, esposito2024}}. 
{Normalization is made using the summation
of the intensities of C-Hoop modes in order to obtain the relative intensities (Int$\rm_{rel}$).}
The computed wavelength and Int$\rm_{rel}$ of the lines are used to {make a spectrum, convolving with a Gaussian}. {The l}ine width of {the full width of half maximum (}FWHM{) of} 10 cm$^{-1}$ is adapted so that individual bands can be seen discretely. We have used QChem (quantum chemistry package) to perform our calculation{s} \citep[]{YSaho15}. {The vibrational transitions can be viewed using a molecular editor/viewing software, for example: IQmol, Gaussview, Avogadro etc.} For our study, molecules are chosen in terms of adjacency of C$-$H units{,} which are then substituted with deuterium to form PAH molecule with solo, duet, trio and quartet C$-$D units.
{Previous observations by \citet[]{Doney15} do not support high deuteration of {PAHs} in the ISM. 
{Considering that, PAHs of small to moderate size with duet, trio and quartet C$-$D is comparatively less likely to exist.}
However, the motivation} of this paper is to see the spectral features arising from C$-$D$\rm_{oop}$ modes of different adjacency type {and therefore, we consider PAHs with solo, duet, trio and quartet C-D unit.}
A molecule can have more than one unique position of solo, duet, trio and quartet C$-$D units and there is no {significant} energy difference depending on the position of deuterium indicating that substitution of D occurs with equal probability in all positions.
{Despite this, we consider deuteration only at a fixed site even if other sites of deuteration are possible to simplify the analysis.} 
Note that a variation in the spectra is expected as we consider different isomers. Also, in the same molecule which exhibits different types of adjacency site simultaneously, for example: pyrene (it has both duet and trio C$-$H/C$-$D sites), there is no preferential site of deuterium substitution for {the} same {number} of substituting D. 

{DFT with harmonic approximation neglects anharmonic effects, such as mode couplings, overtones, and resonances, leading to discrepancies with high-resolution experimental spectra, particularly in the C–H stretching region \citep[e.g.,][]{Mackie15,mackie2016anharmonic}. However, the IR spectral region of our interest here (10$-$20~$\mu$m) is unaffected by anharmonic effects and aligns well with experiments when analyzed using scaled harmonic frequencies \citep[]{lemmens2021infrared}. To examine anharmonic effects in greater detail, we have performed calculations on small PAHs (naphthalene, anthracene, and pyrene) employing a generalized second-order vibrational perturbation theory (GVPT2) at the B3LYP-D3(BJ)/N07D level of theory,
using Gaussian 16 software suite \citep{g16}. This combination significantly improves the accuracy of anharmonic IR spectra
\citep[e.g.,][]{Mackie15, mackie2016anharmonic,mackie2018anharmonic,chen2018carrier,esposito2024,2024MNRAS.533.4150P, Shivani24}. The results were compared with gas-phase spectra from the National Institute of Standards and Technology (NIST) and the scaled harmonic frequencies reported in the present study, as provided in the Appendix section (Figures A1-A3). The comparison shows a good match of the scaled harmonic frequencies with both the anharmonic and NIST data (within 0.8\% of the experiments) in the 10$-$20~$\mu \rm m$ region.
This confirms that in order to study the PAH IR spectra in the 10$-$20~$\mu \rm m$ region, the adapted method using B3LYP/6-311G** and scaling the harmonic frequencies should be sufficient.}
\section{Results}
\begin{figure*}
\includegraphics[width=18cm,height=8cm]{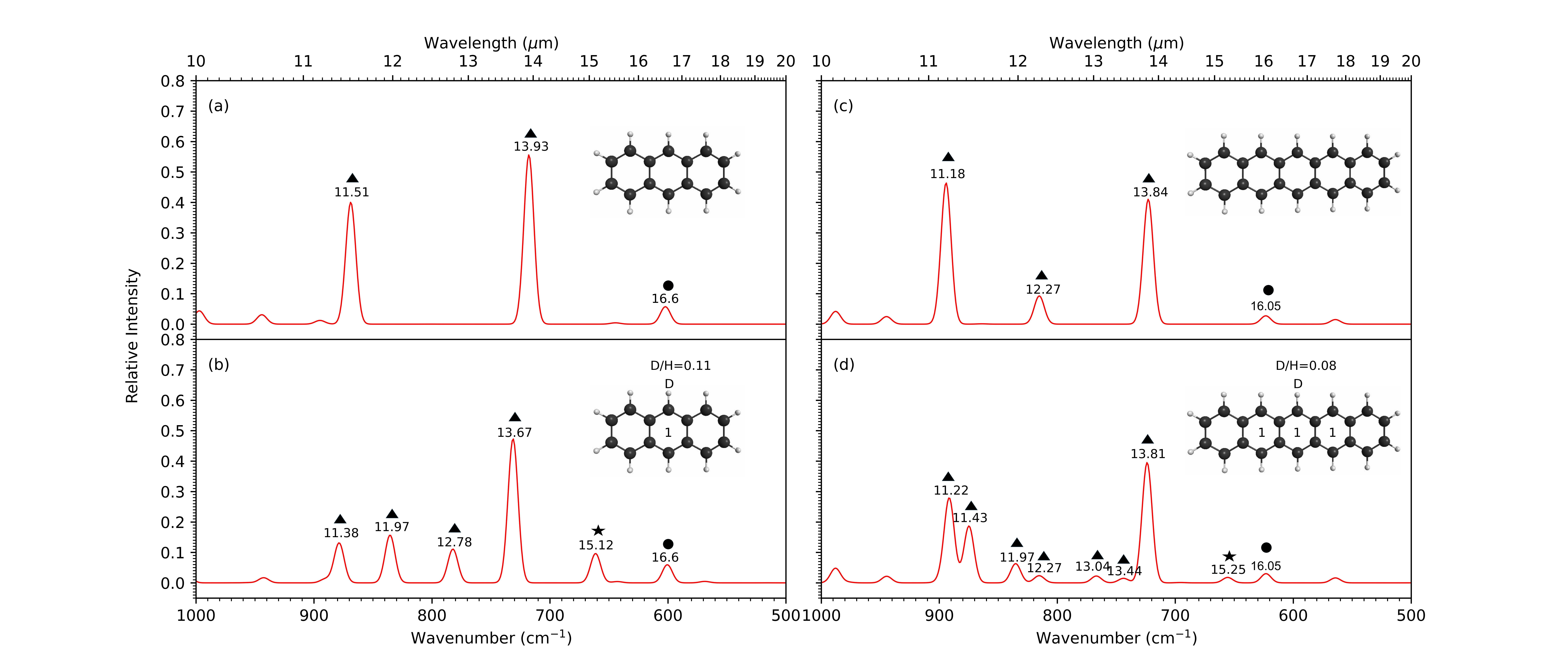}
\caption{{Theoretical spectra of (a) anthracene, (b) singly-deuterated anthracene,
(c) pentacene, (d) singly-deuterated pentacene. The deuterium is placed at a solo C$-$H site.  $\blacktriangle$ represents  C$-$H$\rm_{oop}$ mode, $\medbullet$ represents C$-$C$-$C$\rm_{inplane}$ mode, $\filledstar$ represents C$-$D$\rm_{oop}$ mode.}} 
\label{fig1}
\end{figure*}
In the following sections, we will particularly focus on the C$-$D$\rm_{oop}$ vibrational modes of deuterated PAHs in terms of adjacency of C$-$D bonds and explore {how} deuteration {at the C$-$H site belonging to {a} different adjacency group} affects the existing oop bands and the overall spectra.
\subsection*{PAHs with solo C$-$H group:}
Anthracene (C$_{14}$H$_{10}$) and pentacene (C$_{22}$H$_{14}$) are PAH molecules with both solo and quartet C$-$H sites. Here, we will discuss the characteristic behaviour of deuterated anthracene and deuterated pentacene, when D substitutes H at the solo C$-$H site and later discuss the same molecules,  when D substitution occurs at the quartet C$-$H site {at subsequent steps resulting into solo, duet, trio and quartet C$-$D unit.} {Figure~\ref{fig1} shows the spectra of anthracene and pentacene along with their deuterated counterparts.} {We have marked the relevant significant features
with the calculated wavelengths and also label the peaks with the attributed vibrational modes}. Also, D/H\footnote{$\frac{\rm D}{\rm H}$=$\frac{\rm no.~of~D}{\rm no.~of~H}$} in the PAH species is mentioned in the figure. In a non-deuterated anthracene (Fig~\ref{fig1}a), oop motion of solo C$-$H groups gives a feature at 11.51~$\mu \rm m$ {(Int$\rm_{rel}\sim$~0.4)}. 
This position is comparatively red shifted in comparison to that of the observed emission feature at 11.2~$\mu \rm m$.  
{The oop} vibrational mode of quartet C$-$H groups, which is larger in number{,} gives an intense peak at 13.93~$\mu \rm m$ {(Int$\rm_{rel}\sim$~0.56)}.
When a deuterium {atom replaces} a H {atom} at {a} solo C$-$H site, instead of a single intense peak, two comparatively moderate peaks arise at 11.38 {(Int$\rm_{rel}\sim$~0.13)} and 11.97 $\mu \rm m$ {(Int$\rm_{rel}\sim$~0.16)}
which are characteristics of solo C$-$H$\rm_{oop}$ mode
and {a combination of} both solo and quartet C$-$H$\rm_{oop}$ modes, respectively (Fig~\ref{fig1}b). 
The intense quartet C$-$H$\rm_{oop}$ vibration is slightly blueshifted from 13.93~$\mu \rm m$ to 13.67~$\mu \rm m$ with a decrease in intensity to {Int$\rm_{rel}\sim$~0.47} in singly-deuterated anthracene along with the appearance of a feature at 12.78 $\mu \rm m$ {(Int$\rm_{rel}\sim$~0.11)} due to quartet C$-$H$\rm_{oop}$ modes.
Another unique feature is {at} 15.12~$\mu \rm m$ {(Int$\rm_{rel}\sim$~0.1)} which is characteristic of solo C$-$D$\rm_{oop}$ motion antisymmetric with quartet C$-$H$\rm_{oop}$ motion. The remaining small feature at $\sim$~16.6~$\mu \rm m$ feature (Fig~\ref{fig1}a-b) appear due to C$-$C$-$C$\rm_{inplane}$ bending mode and are not associated with either C$-$H$\rm_{oop}$ or C$-$D$\rm_{oop}$ vibrational modes.

A larger PAH molecule, for example, pentacene (C$_{22}$H$_{14}$){,} shows an intense solo C$-$H$\rm_{oop}$ vibrational mode at 11.18~$\mu \rm m$ {(Int$\rm_{rel}\sim$~0.46)}, a weak antisymmetric solo C$-$H$\rm_{oop}$ mode at 12.27~$\mu \rm m$ {(Int$\rm_{rel}\sim$~0.09)} and a quartet C$-$H$\rm_{oop}$ mode at 13.84~$\mu \rm m$ {(Int$\rm_{rel}\sim$~0.4)} as shown in Fig~\ref{fig1}c. Substituting a H at solo C$-$H site with a D (Fig~\ref{fig1}d) results into two closeby solo C$-$H$\rm_{oop}$ modes appearing at 11.22~$\mu \rm m$ {(Int$\rm_{rel}\sim$~0.28)} and 11.43~$\mu \rm m$ {(Int$\rm_{rel}\sim$~0.19)} and a quartet C$-$H$\rm_{oop}$ mode retaining its position as its non-deutearted counterpart at 13.81~$\mu \rm m$ {(Int$\rm_{rel}\sim$~0.4)}.
The small feature at 12.27~$\mu \rm m$ in non-deuterated pentacene continues to be present with a much reduced intensity alongside an additional {C$-$H$\rm_{oop}$ mode} at 11.97~$\mu \rm m$ {(Int$\rm_{rel}\sim$~0.06)} in singly-deuterated pentance (Figure~\ref{fig1}d). Deuteration also produces two minor C$-$H$\rm_{oop}$ modes at 13.04~$\mu \rm m$ {(Int$\rm_{rel}\sim$~0.02)} and 13.44~$\mu \rm m$ {(Int$\rm_{rel}\sim$~0.02)} in Figure~\ref{fig1}d. 
The solo C$-$D$\rm_{oop}$ vibrational mode, which is apparently the direct manifestation of deuteration of PAHs, is vaguely present at 15.25~$\mu \rm m$ with {Int$\rm_{rel}\sim$~0.02} (Fig~\ref{fig1}d) in singly-deuterated pentance.  The two minor features at $\sim$16.05~$\mu \rm m$ and $\sim$17.72~$\mu \rm m$ are present in the spectra of both the forms of pentacene and are due to C$-$C$-$C$\rm_{inplane}$ bending modes. {Note that we have marked the wavelengths (in all spectra) which are relevant for the discussion of this paper.} 
\subsection*{PAHs with duet C$-$H group:}
\begin{figure*}
\includegraphics[width=18cm,height=14cm]{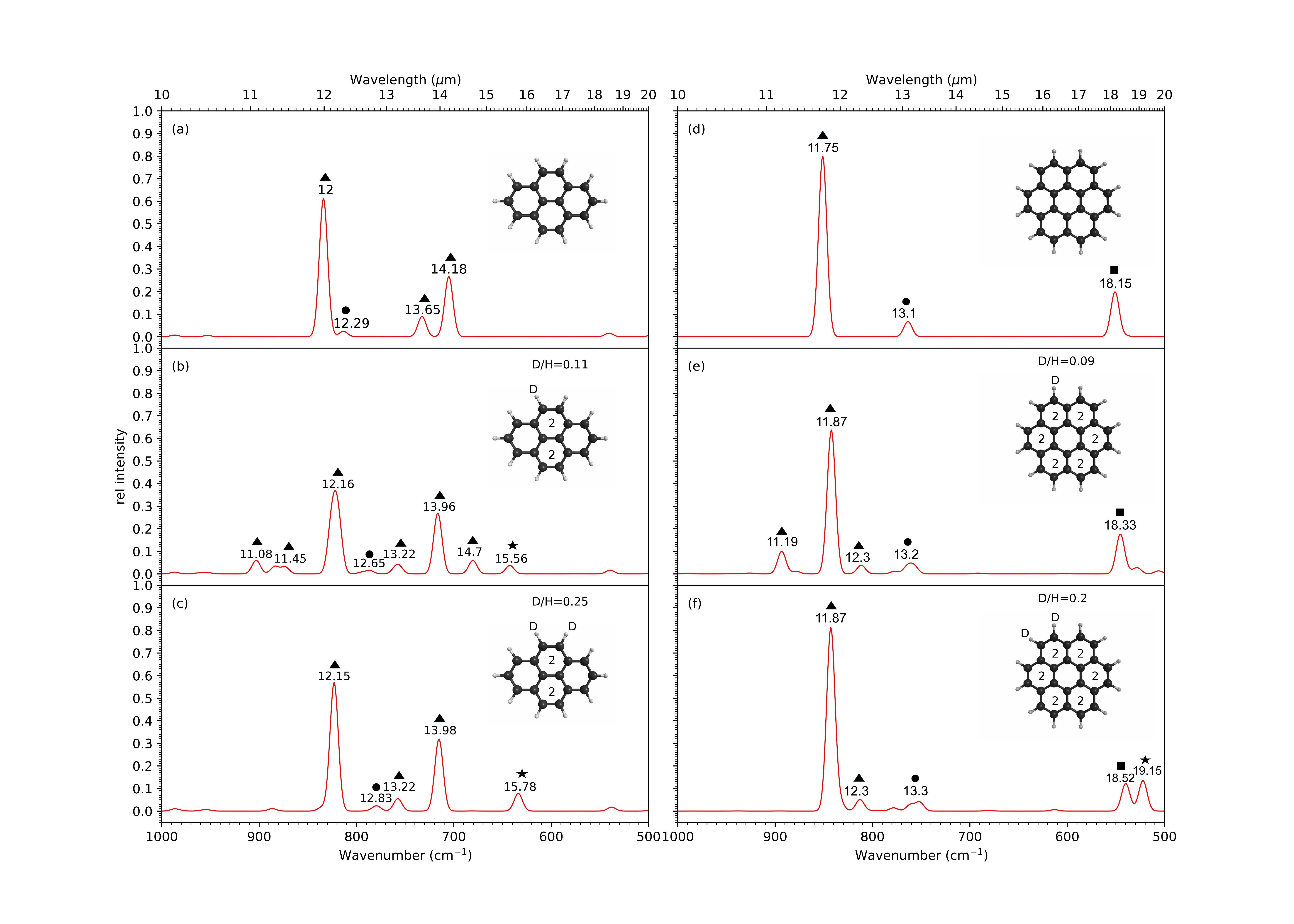}
\vspace{-3em}
\caption{{Theoretical spectra of (a) pyrene, (b) singly-deuterated pyrene, (c) doubly-deuterated pyrene,
(d) coronene, (e) singly-deuterated coronene, (f) doubly-deuterated coronene. The deuterium is {placed at} duet C$-$H site. {$\blacktriangle$ represents  C$-$H$\rm_{oop}$ mode, $\medbullet$ represents C$-$C$-$C$\rm_{inplane}$ mode, $\blacksquare$ represents blending mode of C$-$C$-$C$\rm_{oop}$ and C$-$H$\rm_{oop}$ vibrations, $\filledstar$ represents C$-$D$\rm_{oop}$ mode.}}}
\label{fig2}
\end{figure*}
Pyrene (C$_{16}$H$_{10}$) has both duet and trio C$-$H sites, at which we can substitute H with D. In this section, we discuss vibrational features of deuterated pyrene when D substitutes H at the duet C$-$H site. A single D substitution will first give rise to a solo C$-$H and a solo C$-$D bond at the substitution site, while two {substituting} Ds will give rise to a duet C$-$D unit. Since, there is no preferential site for {deuteration}, the placement of second deuterium can occur at any C$-$H site (either duet or trio C$-$H unit or in a different benzene ring), but we constrain ourselves to place the deuterium atoms at the duet C$-$H site in the same ring to generate a duet C$-$D unit. In case of non-deuterated neutral pyrene, the strong peak near 12~$\mu \rm m$ {(Int$\rm_{rel}\sim$~0.61)} and 14.18~$\mu \rm m$ {(Int$\rm_{rel}\sim$~0.27)} with {another moderately intense peak} at 13.65~$\mu \rm m$ {(Int$\rm_{rel}\sim$~0.09)} are manifestations of duet and trio C$-$H$\rm_{oop}$ vibrational modes (Figure~\ref{fig2}a). {The remaining much weaker features are due to C$-$C$-$C$\rm_{inplane}$ bending}. On substituting H at the duet C$-$H site with a single and double deuterium {atoms}, there is a small change in wavelength positions by 0.1 to 0.2~$\mu \rm m$ and intensities of these bands along with a few new features emerging (Figure~\ref{fig2}b-c). The {feature at} 13.65~$\mu \rm m$ blueshifts to 13.22~$\mu \rm m$ {with a reduced intensity} in case of both singly and doubly-deuterated pyrene and a new feature due to C$-$H$\rm_{oop}$ vibration appears at 14.7~$\mu \rm m$ {(Int$\rm_{rel}\sim$~0.06)} only in case of singly-deuterated pyrene. The C$-$C$-$C$\rm_{inplane}$ bending feature at 12.29~$\mu \rm m$ {(Int$\rm_{rel}\sim$~0.02)} in non-deuterated pyrene shifts to 12.58$-$12.72~$\mu \rm m$ {(Int$\rm_{rel}\sim$~0.01)} in singly-deuterated pyrene and to 12.83~$\mu \rm m$ {(Int$\rm_{rel}\sim$~0.02)} in doubly-deuterated pyrene.
New features include a few modes at 11.08~$\mu \rm m$ {(Int$\rm_{rel}\sim$~0.05)}, 11.31~$\mu \rm m$ {(Int$\rm_{rel}\sim$~0.03)} and 11.45~$\mu \rm m$ {(Int$\rm_{rel}\sim$~0.03)} that emerge in singly-deuterated pyrene. Unlike non-deuterated pyrene that lacks a solo C$-$H unit and does not show a feature at $\sim$~11.2~$\mu \rm m$, substituting by a single D at one of the two available duet C$-$H site{s} enables the pyrene molecule to {gain a solo C$-$H unit and produces these solo C$-$H$\rm_{oop}$ modes {nearly at $\sim$~11.1~$\mu \rm m$.}
There also exists a C$-$D$\rm_{oop}$ vibrational mode at 15.56~$\mu \rm m$ {(Int$\rm_{rel}\sim$~0.04)} for 
singly-deuterated and 15.78~$\mu \rm m$ {(Int$\rm_{rel}$~0.08)} for a doubly-deuterated pyrene.
\begin{figure*}
\includegraphics[width=18cm,height=14cm]{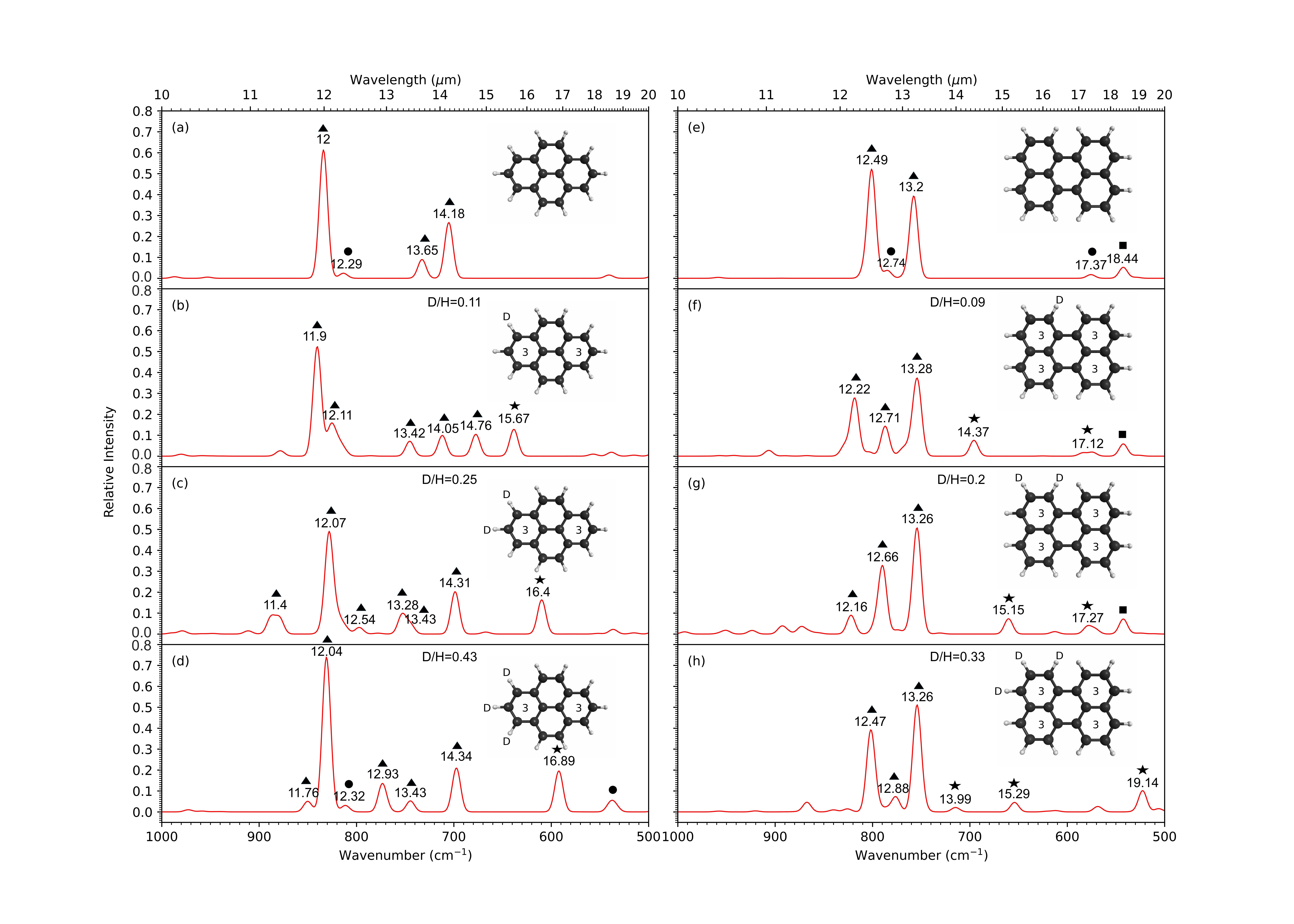}
\vspace{-3em}
\caption{{Theoretical spectra of (a) pyrene, (b) singly-deuterated pyrene, (c) doubly-deuterated pyrene, (d) triply-deuterated pyrene,
(e) perylene, (f) singly-deuterated perylene, (g) doubly-deuterated perylene, (h) triply-deuterated perylene. The location of deuterium substitution is {fixed at} a trio C$-$H site. {$\blacktriangle$ represents  C$-$H$\rm_{oop}$ mode, $\medbullet$ represents C$-$C$-$C$\rm_{inplane}$ mode, $\blacksquare$ represents blending mode of C$-$C$-$C$\rm_{oop}$ and C$-$H$\rm_{oop}$ vibrations, $\filledstar$ represents C$-$D$\rm_{oop}$ mode.}}}
\label{fig3}
\end{figure*}

Coronene (C$_{24}$H$_{12}$) is a compact, symmetric molecule with all the peripheral C$-$H sites being duet or doubly-adjacent and {the corresponding spectra is ordinary compared to other PAHs (Fig~\ref{fig2}d).} The distinct features of a non-deuterated coronene in the range of 10$-$20~$\mu \rm m$ are duet C$-$H$\rm_{oop}$ modes at 11.75~$\mu \rm m$ {(Int$\rm_{rel}\sim$~0.8)} and C$-$C$-$C$\rm_{oop}$ modes (partially contributed by duet C$-$H$\rm_{oop}$ modes) at 18.15~$\mu \rm m$ {(Int$\rm_{rel}\sim$~0.2)}. 
The small features {near} $\sim$~13.1~$\mu \rm m$ {(Int$\rm_{rel}\sim$~0.07)} are due to C$-$C$-$C$\rm_{inplane}$ bending modes. On deuteration, {duet} C$-$H$\rm_{oop}$ mode {at 11.75~$\mu \rm m$ is} slightly shifted to
11.87~$\mu \rm m$ in both singly and doubly-deuterated coronene with {Int$\rm_{rel}\sim$~0.64 and $\sim$~0.81}, respectively (Figure~\ref{fig2}e-f). Deuteration also produces additional C$-$H$\rm_{oop}$ features at 12.3~$\mu \rm m$ {(Int$\rm_{rel}\sim$~0.04-0.05)} in singly and doubly-deuterated coronene.
The 18.15~$\mu \rm m$ in non-deuterated coronene shifts to 18.33~$\mu \rm m$ in singly-deuterated coronene with additional 
features of smaller intensities in the longward side. The same feature is present at 18.52~$\mu \rm m$ in doubly-deuterated coronene. A {feature due to overlapping} of C$-$D$\rm_{oop}$ and C$-$C$-$C$\rm_{oop}$ modes
is present at 19.15~$\mu \rm m$ (Int$\rm_{rel}\sim$~0.13) in doubly-deuterated coronene (Fig~\ref{fig2}f). Besides, deuterium substitution produces additional vibrational modes at $\sim$~11.19~$\mu \rm m$ {(Int$\rm_{rel}\sim$~0.1)} in singly-deuterated coronene for a similar reason as in singly-deuterated pyrene due to solo C$-$H$\rm_{oop}$ {mode} (Figure~\ref{fig2}e). 
\begin{figure*}
\hspace{-4em}
\includegraphics[width=18cm,height=14cm]{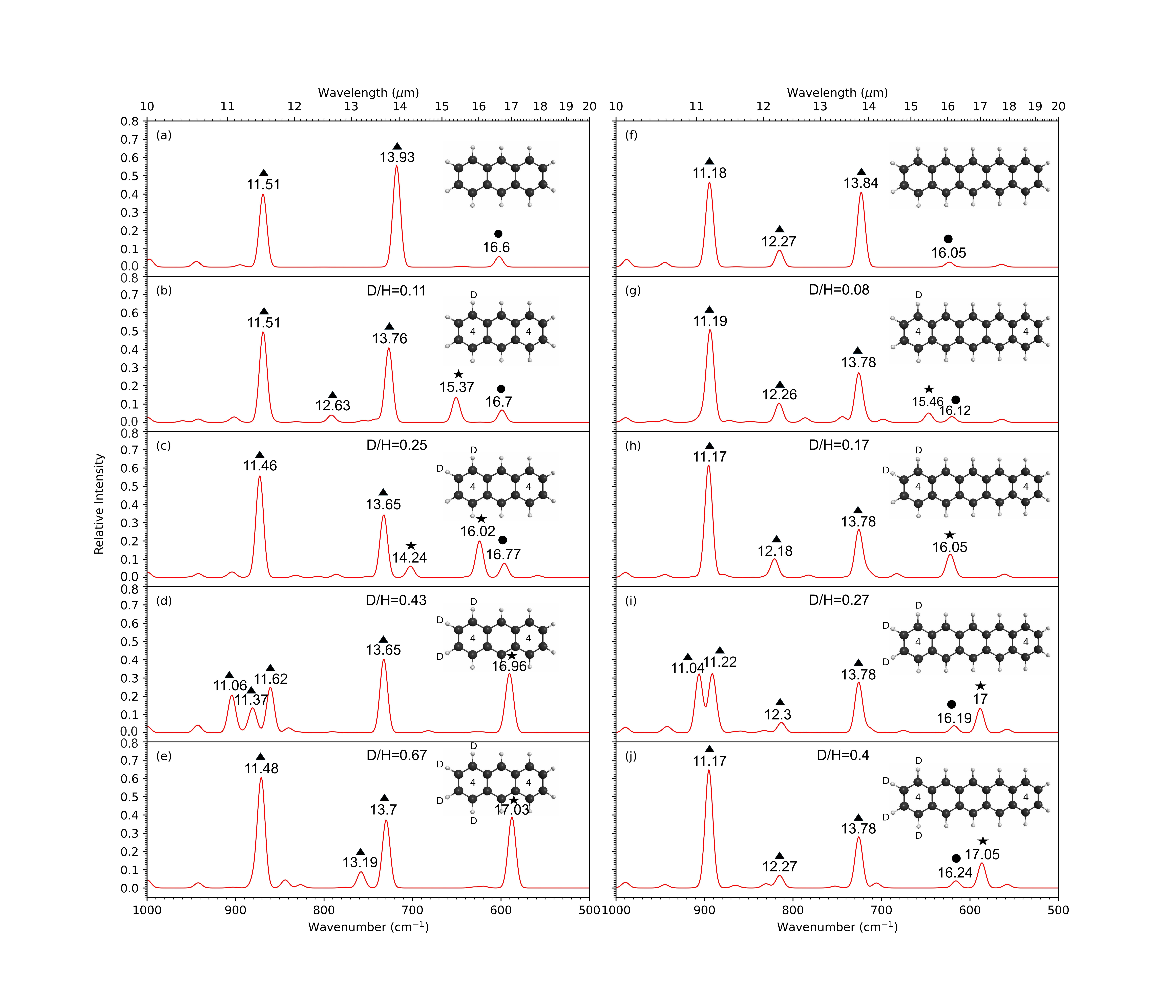}
\vspace{-3em}
\caption{{Theoretical spectra of (a) anthracene, (b) singly-deuterated anthracene, (c) doubly-deuterated anthracene, (d) triply-deuterated anthracene,
(e) quadruply-deuterated anthracene, (f) pentacene , (g) singly-deuterated pentacene, (h) doubly-deuterated pentacene, (i) triply-deuterated pentacene \& (j) quadruply-deuterated pentacene. The location of deuterium substitution is {fixed at} a quartet C$-$H site. 
{$\blacktriangle$ represents  C$-$H$\rm_{oop}$ mode, $\medbullet$ represents C$-$C$-$C$\rm_{inplane}$ mode, $\filledstar$ represents C$-$D$\rm_{oop}$ mode.}}}
\label{fig4}
\end{figure*}
\subsection*{PAHs with trio C$-$H group:}
As mentioned in previous section, pyrene has trio C$-$H sites{,} too{,} apart from duet C$-$H sites and this section 
will discuss C$-$H$\rm_{oop}$ vibrational features, when single or multiple Ds substitute H atoms at the trio C$-$H sites of pyrene. The subsequent D substitutions at trio C$-$H sites {in} the same benzene ring will produce solo, duet and trio C$-$D units {at} subsequent steps.
{If we start adding deuterium (starting from {number} of D= 1 to 3) at subsequent steps, single deuterium substitution will give rise to a solo C-D unit. When a second deuterium is added, a duet C-D unit will be created. In this way, placing a third  deuterium will give rise to a trio C-D unit.}
The calculated spectra are shown in Fig~\ref{fig3}(a-d). The features of non-deuterated pyrene have already been discussed in previous section. A clear difference can be seen between the spectral features of a deuterated pyrene depending upon whether D is {placed at}  a duet or {a} trio C$-$H site (Figure~\ref{fig2} and \ref{fig3}){,} 
indicating that the position of D affects the spectra. Unlike non-deuterated pyrene that shows two distinct features between 13$-$15~$\mu \rm m$, deuteration at 
trio C$-$H sites would arise several features, which are majorly C$-$H$\rm_{oop}$ vibrational modes (Figure~\ref{fig3}b-d). The comparatively intense feature at $\sim$~12~$\mu \rm m$ in non-deuterated pyrene almost appears at {a} similar wavelength upon deuteration with a slight reduction in intensity (Figure~\ref{fig3}b-d), {giving off neighbouring features due to either more C$-$H$\rm_{oop}$ modes or C$-$C$-$C$\rm_{inplane}$ modes.} 
{The feature at 13.65~$\mu \rm m$ in non-deuterated pyrene shifts to $\sim$13.42-13.43~$\mu \rm m$ in singly, doubly and triply-deuterated pyrene.} The presence of features at 11.3~$\mu \rm m$ {(Int$\rm_{rel}\sim$~0.08)} and 11.4~$\mu \rm m$ {(Int$\rm_{rel}\sim$~0.08)} (Figure~\ref{fig3}c) is clearly visible arising from C$-$H$\rm_{oop}$ modes of solo C$-$H unit, which is created when {two deuterium atoms replace two H atoms at the trio C$-$H site.}
In addition, C$-$H$\rm_{oop}$ modes combined with C$-$D$\rm_{oop}$ modes contribute to produce features at {14.76~$\mu \rm m$ {(Int$\rm_{rel}\sim$~0.1)} in a singly-deuterated pyrene,}
at 13.28~$\mu \rm m$ {(Int$\rm_{rel}\sim$~0.1)} and 14.31~$\mu \rm m$ {(Int$\rm_{rel}\sim$~0.2)} in a doubly-deuterated pyrene (Fig~\ref{fig3}c) and at 12.93~$\mu \rm m$ {(Int$\rm_{rel}$~0.14)} and 14.34~$\mu \rm m$ {(Int$\rm_{rel}\sim$~0.21)} in a triply-deuterated pyrene  (Fig~\ref{fig3}d).
A feature majorly due to {a} {solo} C$-$D$\rm_{oop}$ mode is present at 15.67~$\mu \rm m$ {(Int$\rm_{rel}\sim$~0.13}, Fig~\ref{fig3}b) in singly-deuterated pyrene (a single
deuterium will correspond to solo C$-$D$\rm_{oop}$ mode). The analogous duet and trio C$-$D$\rm_{oop}$ modes appear at a longer wavelength position of 16.4~$\mu \rm m$  {(Int$\rm_{rel}\sim$~0.16}, figure~\ref{fig3}c) for a doubly-deuterated pyrene and 16.89~$\mu \rm m$  {(Int$\rm_{rel}\sim$~0.2}, figure~\ref{fig3}d) for a triply-deuterated pyrene, respectively. 

Another example of PAH molecule with trio C$-$H site is perylene (C$_{20}$H$_{12}$) and {F}igure~\ref{fig3}(e-h) shows the calculated spectra for non-deuterated perylene (C$_{20}$H$_{12}$) and deuterated perylenes with number of D=1, 2 \& 3, where D substitutes H at the trio C$-$H site. The most important features in a non-deuterated perylene in the $11-15~\mu \rm m$ are trio C$-$H$\rm_{oop}$ modes that occur at 12.49~$\mu \rm m$ {(Int$\rm_{rel}\sim$~0.52)} and 13.2~$\mu \rm m$ {(Int$\rm_{rel}\sim$~0.39)} as shown in Fig~\ref{fig3}e. There exists a less intense C$-$C$-$C$\rm_{inplane}$ bending mode at 12.74~$\mu \rm m$ {(Int$\rm_{rel}\sim$~0.04)}. Remaining small features at 17.37~$\mu \rm m$ and 18.44~$\mu \rm m$ are due to C$-$C$-$C$\rm_{inplane}$ bending and {blending mode of C$-$C$-$C$\rm_{oop}$ and C$-$H$\rm_{oop}$ vibrations}, respectively. {Similar to previous scenarios,} on account of adding single or multiple Ds at the trio C$-$H site, these features slightly change {the} wavelength positions along with the appearance of new small features, which are mostly C$-$H$\rm_{oop}$ modes or other C$-$H/C$-$C$-$C bond vibration{s}. An 
effect of deuterium substitution can be seen at 14.37~$\mu \rm m$ {(Int$\rm_{rel}\sim$~0.08)} and 17.12~$\mu \rm m$ {(Int$\rm_{rel}\sim$~0.02)} in a singly-deuterated perylene (Fig~\ref{fig3}f), where C$-$D$\rm_{oop}$ {mode}, {overlapping} with other modes
is present. Similar modes appear at 15.15~$\mu \rm m$ {(Int$\rm_{rel}\sim$~0.07)}
and 17.27~$\mu \rm m$ {(Int$\rm_{rel}\sim$~0.04)} for doubly-deuterated perylene and at 13.99~$\mu \rm m$ {(Int$\rm_{rel}\sim$~0.02)}, 15.29~$\mu \rm m$ {(Int$\rm_{rel}\sim$~0.05)} and 19.14~$\mu \rm m$ {(Int$\rm_{rel}\sim$~0.1)} for triply-deuterated perylene.
\subsection*{PAHs with quartet C$-$H group:}
In this section, we present the spectra of deuterated anthracene and deuterated pentacene  when deuterium substitution occurs at quartet C$-$H site.
The characteristics of the non-deuterated anthracene and non-deuterated pentacene are described 
before. Deuterium substitution at the quartet C$-$H site does not {significantly} change the position of the solo C$-$H$\rm_{oop}$ vibrational modes in deuterated anthracenes (Fig~\ref{fig4}b-e). 
Exception is a triply-deuterated anthracene (Fig~\ref{fig4}d), in which an additional solo C$-$H site is created besides the existing solo C$-$H sites because of deuterium substitution, which results into several solo C$-$H$\rm_{oop}$ modes. These modes occur at 11.06, 11.37, {and} 11.62~$\mu \rm m$.
The other intense peak at 13.93~$\mu \rm m$ due to quartet C$-$H$\rm_{oop}$ modes in non-deuterated anthracene 
appears at the range of 13.7-13.8~$\mu \rm m$ in deuterated anthracenes with a decrease in intensity. {An additional weak feature due to C$-$H$\rm_{oop}$ modes appears at 12.63~$\mu \rm m$ {(Int$\rm_{rel\sim}$~0.04)} in singly-deuterated anthracene (Figure~\ref{fig4}b)}.
The effect of deuterium substitution is seen near 15.37~$\mu \rm m$ {(Int$\rm_{rel}$~0.14)} (Fig~\ref{fig4}b) which is a C$-$D$\rm_{oop}$ {vibrational mode} merged with C$-$H$\rm_{oop}$ mode.
When we deuterate two of the four quartet C$-$H sites in the same ring, the same feature (antisymmetric and symmetric) 
appears at 14.24~$\mu \rm m$ {(Int$\rm_{rel}$~0.06)} and 16.02~$\mu \rm m$ {(Int$\rm_{rel}$~0.2)} respectively (Fig~\ref{fig4}c). On further increasing the deuteration upto $D=3$ {and} 4 in the same ring, only the
C$-$D$\rm_{oop}$ {mode} exists eliminating the C$-$H$\rm_{oop}$ modes from merging and is redshifted
to 16.96~$\mu \rm m$ {( Int$\rm_{rel}$~0.32}, Fig~\ref{fig4}d) and 17.03~$\mu \rm m$ ( {Int$\rm_{rel}$~0.39}, Fig~\ref{fig4}e) respectively. In addition, for quadruply-deuterated anthracene, there is an excess mode of quartet C$-$H$\rm_{oop}$ with contribution from quartet C$-$D$\rm_{oop}$ modes
at 13.19~$\mu \rm m$ {( Int$\rm_{rel}$~0.09}, Fig~\ref{fig4}e). The 16.6~$\mu \rm m$ feature due to C$-$C$-$C$\rm_{inplane}$ bending in non-deuterated anthracene persists to occur at {a} similar wavelength after deuteration, although, with increasing deuteration, its position overlaps with that of C$-$D$\rm_{oop}$ modes and cannot be distinguished.

Fig~\ref{fig4}(f-j) presents the computational vibrational spectra of pentacene along with its deuterated counterparts with no. of D=1, 2, 3 {and} 4, when D {replaces H} at the quartet C$-$H sites, resulting into solo, duet, trio and quartet C$-$D units at subsequent steps\footnote{{This is {the} same as the creation of solo, duet and trio C-D units in a deuterated pyrene, previously explained.}.}.
The solo C$-$H$\rm_{oop}$ modes at $\sim$~11.18~$\mu \rm m$ remain invariant upon deuteration.
Exception is a triply-deuterated pentacene (Figure~\ref{fig4}i), in which triple deuteration in the same ring at the edge creates a solo C$-$H unit, the oop vibration of which produces a feature at 11.04~$\mu \rm m$ {(Int$\rm_{rel}$~0.32)}, next to the 11.22~$\mu \rm m$ {with} similar Int$\rm_{rel}$.
The quartet C$-$H$\rm_{oop}$ mode appears {at} 13.78~$\mu \rm m$ upon deuteraion, close to the position as in a non-deuterated pentacene, with a {reduction in relative intensity. A few new features appear mostly due to  C$-$H$\rm_{oop}$ modes and these are weak in intensity.} A C$-$D$\rm_{oop}$ mode arises which tend{s} to be redshifted with {the} increasing 
number of deuterium ({no. of} D=1,2,3,4) and appears at 15.46~$\mu \rm m$ {(Int$\rm_{rel}$~0.05)}, 16.05~$\mu \rm m$ {(Int$\rm_{rel}$~0.13)},
17~$\mu \rm m$ {(Int$\rm_{rel}$~0.13)} and 17.05~$\mu \rm m$ {(Int$\rm_{rel}$~0.13)}{,} respectively{,} as shown in  Fig~\ref{fig4}(g-j). 
\begin{figure*}
\hspace{-4em}
\includegraphics[width=18cm,height=10cm]{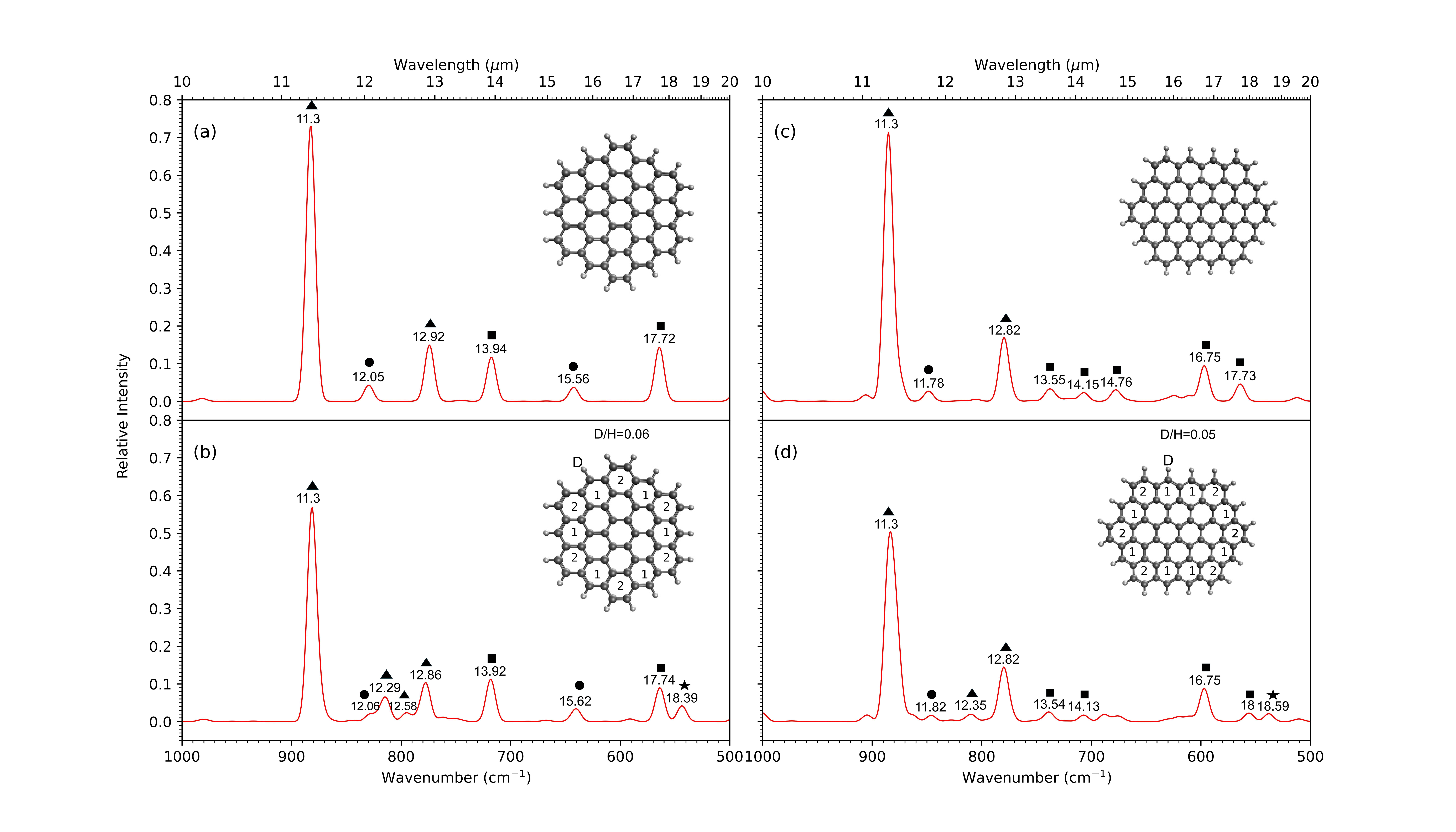}
\caption{{Theoretical spectra of (a) circumcoronene, (b) singly-deuterated circumcoronene, (c) circumovalene, (d) singly-deuterated circumovalene. The location of deuterium substitution is {fixed at} a solo C$-$H site. {$\blacktriangle$ represents  C$-$H$\rm_{oop}$ mode, $\medbullet$ represents C$-$C$-$C$\rm_{inplane}$ mode, $\blacksquare$ represents blending mode of C$-$C$-$C$\rm_{oop}$ and C$-$H$\rm_{oop}$ vibrations, $\filledstar$ represents C$-$D$\rm_{oop}$ mode.}}}
\label{fig5}
\end{figure*}
\begin{figure*}
\hspace{-4em}
\includegraphics[width=18cm,height=14cm]{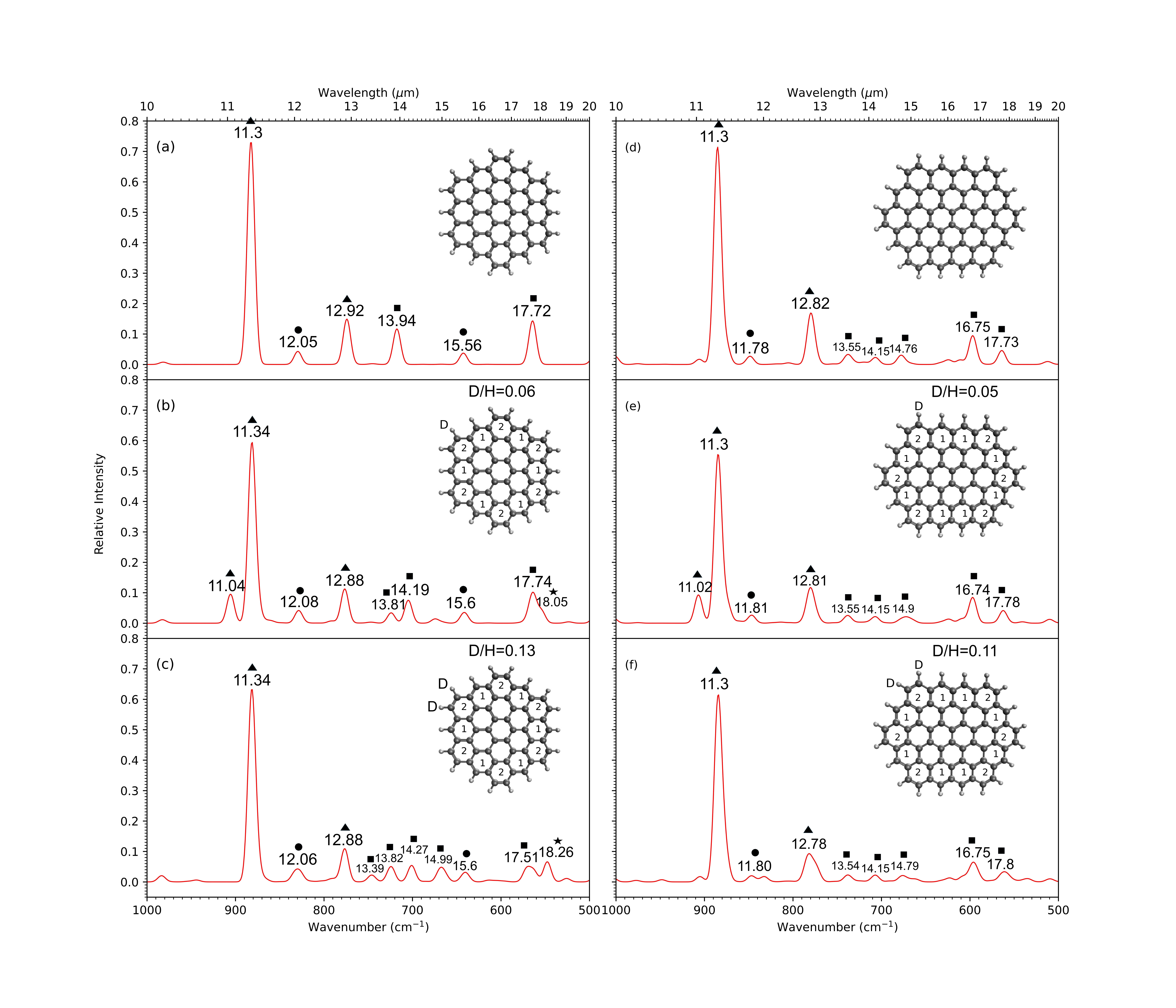}
\vspace{-2em}
\caption{{Theoretical spectra of (a) circumcoronene, (b) singly-deuterated circumcoronene, (c) doubly-deuterated circumcoronene,
(d) circumovalene, (e) singly-deuterated circumovalene, (f) doubly-deuterated circumovalene. The location of deuterium substitution is {fixed at} a duet C$-$H site. {$\blacktriangle$ represents  C$-$H$\rm_{oop}$ mode, $\medbullet$ represents C$-$C$-$C$\rm_{inplane}$ mode, $\blacksquare$ represents blending mode of C$-$C$-$C$\rm_{oop}$ and C$-$H$\rm_{oop}$ vibrations, $\filledstar$ represents C$-$D$\rm_{oop}$ mode.}}}
\label{fig6}
\end{figure*}
\subsection*{Large compact PAHs with deuterium:}
This paper also reports C$-$D$\rm_{oop}$ features from large compact PAHs, which are expected to survive the extreme condition of the interstellar environment more readily. As PAHs grow into {a} compact and large form, for example: circumcoronene {(C$_{54}$H$_{18}$)} and circumovalene {(C$_{66}$H$_{20}$)}, they would possess more solo and duet C$-$H sites and lack trio and quartet C$-$H sites. {Figure~\ref{fig5} represents the spectra of circumcoronene and cicrcumovalene along with their deuterated counterparts when D substitutes H at the solo C$-$H site. In a non-deuterated circumcoronene (Figure~\ref{fig5}a), the intense peak at 11.3~$\mu \rm m$ {(Int$\rm_{rel}\sim$~0.73)} arises majorly due to {a} solo C$-$H$\rm_{oop}$ vibrational mode. The remaining small peaks in the {10$-$20~$\mu \rm m$} range are at {12.05~$\mu \rm m$ (Int$\rm_{rel}\sim$~0.04) and 15.56~$\mu \rm m$ (Int$\rm_{rel}\sim$~0.04) due to C$-$C$-$C$\rm_{inplane}$ bending, 12.92~$\mu \rm m$ (Int$\rm_{rel}\sim$~0.15) due to an antisymmetric oop motion of solo and duet C$-$H groups, 13.94~$\mu \rm m$ (Int$\rm_{rel}\sim$~0.12) and 17.72~$\mu \rm m$ (Int$\rm_{rel}\sim$~0.14) due to blending mode of C$-$C$-$C$\rm_{oop}$ and C$-$H$\rm_{oop}$ vibrations}.
On deuteration of a solo C$-$H group, the intensity of {the} solo C$-$H$\rm_{oop}$ mode at 11.3~$\mu \rm m$ is slightly reduced to {Int$\rm_{rel}\sim$~0.56}, without changing the position and part of the intensity is redistributed to give a small feature at 12.29~$\mu \rm m$ {(Int$\rm_{rel}\sim$~0.06)} {and 12.58~$\mu \rm m$ {(Int$\rm_{rel}\sim$~0.02)} etc., majorly due to {a} duet C$-$H$\rm_{oop}$ and some C$-$D$\rm_{oop}$ mode{s} (Figure~\ref{fig5}b), apart from the features that are common to its non-deuterated counterpart. {A {solo} C$-$D$\rm_{oop}$ mode partially merging with C$-$H$\rm_{oop}$ modes is present at 18.39~$\mu \rm m$ {(Int$\rm_{rel}\sim$~0.04).}
For larger PAHs, say, circumovalene (Figure~\ref{fig5}c-d), similar behaviour is seen, when we move from non-deuterated to deuterated PAH. The energy of solo C$-$H$\rm_{oop}$ at {a} $\sim$11.3~$\mu \rm m$ is slightly reduced, which gets distributed to neighboring modes upon deuteration. {A {solo} C$-$D$\rm_{oop}$ mode combined with {other modes}
is observed at 18.59~$\mu \rm m$ {(Int$\rm_{rel}\sim$~0.02)} in singly-deuterated circumovalene. Remaining new features are mostly not contributed by the substituting C$-$D bond.}}

Both circumcorone and circumovalene carry duet C$-$H groups and {their} spectra are shown for these species {in} Figure~\ref{fig6}, when D substitutes H at the duet site at subsequent steps, producing a solo C$-$D unit first and then a duet C$-$D unit.
Similar to all previous scenarios, 
less intense features arise besides the already existing features 
in case of both circumcoronene and circumovalene (Figure~\ref{fig6}a-f).
Since Ds are at the duet site, it does not change the intensity and position of solo C$-$H$\rm_{oop}$ mode at 11.3~$\mu \rm m$. An additional solo C$-$H unit is created because of single deuterium substitution at the duet C$-$H site of circumcoronene and circumovalene, {resulting} {an} additional solo C$-$H$\rm_{oop}$ mode at 11.04~$\mu \rm m$ {(Int$\rm_{rel}\sim$~0.09)} and at 11.02~$\mu \rm m$ {(Int$\rm_{rel}\sim$~0.09)}, respectively in singly-deuterated circumcoronene and singly-deuterated circumovalene.
D substitution should somewhat affect the duet C$-$H$\rm_{oop}$ modes. However, duet C$-$H$\rm_{oop}$ modes are not the strongest in case of circumcoronene as well as circumovalene and any change in them is insignificant apart from giving off additional features {in this region}. A C$-$D$\rm_{oop}$ mode combined with {other modes}
is present at 18.05~$\mu \rm m$ {(Int$\rm_{rel}\sim$~0.04)} and at 18.26~$\mu \rm m$ {(Int$\rm_{rel}\sim$~0.07)} for singly and doubly-deuterated circumcoronene, respectively. 
The C$-$D$\rm_{oop}$ modes in deuterated circumovalenes, however, are overpowered by other modes{,} making it less apparent. 
\section{Discussion:}
\begin{table*}
\centering
\caption{Band position and relative intensity of the existing solo C$-$H$\rm_{oop}$ modes upon deuteration}
\label{tab1}
\newcommand{\mc}[3]{\multicolumn{#1}{#2}{#3}}
\begin{tabular}[c]{c|c|c}
\hline
\hline
 PAH & $\lambda$ (Int$\rm_{rel}$) & $\lambda$ (Int$\rm_{rel}$) \\ 
  & for non-deuterated & for singly-deuterated  \\ \hline
  Anthracene & 11.51~$\mu \rm m$ {(0.4)} & 11.38~$\mu \rm m$ {(0.13)}  \\
    &  & 11.97~$\mu \rm m$ {(0.16)} \\ \hline
    Pentacene & 11.18~$\mu \rm m$ {(0.46)} & 11.22~$\mu \rm m$ {(0.28)} \\
    & & 11.43~$\mu \rm m$ {(0.19)}  \\ \hline
    Circumcoronene & 11.3~$\mu \rm m$ {(0.73)} & 11.3~$\mu \rm m$ {(0.56)} \\ \hline
    Circumovalene & 11.3~$\mu \rm m$ {(0.71)} & 11.3~$\mu \rm m$ {(0.5)}  \\ \hline  
\end{tabular}
\end{table*}
\begin{table*}
\centering
\caption{Band position and relative intensity of the existing duet C$-$H$\rm_{oop}$ modes upon deuteration}
\label{tab2}
\newcommand{\mc}[3]{\multicolumn{#1}{#2}{#3}}
\begin{tabular}[c]{c|c|c|c}
\hline
\hline
 PAH & $\lambda$ (Int$\rm_{rel}$) & $\lambda$ (Int$\rm_{rel}$) & $\lambda$ (Int$\rm_{rel}$)  \\ 
  & for non-deuterated & for singly-deuterated & for doubly-deuterated  \\ \hline
  Pyrene & 12~$\mu \rm m$ {(0.61)} & 12.16~$\mu \rm m$ {(0.37)} & 12.15~$\mu \rm m$ {(0.57)} \\ \hline
    Coronene & 11.75~$\mu \rm m$ {(0.8)} & 11.87~$\mu \rm m$ {(0.64)} & 11.87~$\mu \rm m$ {(0.81)}  \\ \hline
    Circumcoronene & 12.92~$\mu \rm m$ {(0.15)} & 12.88~$\mu \rm m$ {(0.11)}  & 12.88~$\mu \rm m$ {(0.11)} \\ \hline
    Circumovalene & 12.82~$\mu \rm m$ {(0.16)} & 12.81~$\mu \rm m$ {(0.11)} & 12.78~$\mu \rm m$ {(0.09)} \\ \hline  
\end{tabular}
\end{table*}
\begin{table*}
\centering
\caption{Band position and relative intensity of the existing trio C$-$H$\rm_{oop}$ modes upon deuteration}
\label{tab3}
\newcommand{\mc}[3]{\multicolumn{#1}{#2}{#3}}
\begin{tabular}[c]{c|c|c|c|c}
\hline
\hline
 PAH & $\lambda$ (Int$\rm_{rel}$) & $\lambda$ (Int$\rm_{rel}$) & $\lambda$ (Int$\rm_{rel}$) & $\lambda$ (Int$\rm_{rel}$)  \\ 
  & for non-deuterated & for singly-deuterated & for doubly-deuterated & for triply-deuterated \\ \hline
  Perylene & 12.49~$\mu \rm m$ {(0.51)} & 12.22~$\mu \rm m$ {(0.28)} & 12.66~$\mu \rm m$ {(0.33)} & 12.47 {(0.39)} \\  
  & 13.2~$\mu \rm m$ {(0.39)} & 13.28~$\mu \rm m$ {(0.38)} & 13.26~$\mu \rm m$ {(0.51)} & 13.26 {(0.51)} \\ \hline
\end{tabular}
\end{table*}
\begin{table*}
\centering
\caption{Band position and relative intensity of the existing quartet C$-$H$\rm_{oop}$ modes upon deuteration}
\label{tab4}
\footnotesize
\newcommand{\mc}[3]{\multicolumn{#1}{#2}{#3}}
\begin{tabular}[c]{c|c|c|c|c|c}
\hline
\hline
 PAH & $\lambda$ (Int$\rm_{rel}$) & $\lambda$ (Int$\rm_{rel}$) & $\lambda$ (Int$\rm_{rel}$) & $\lambda$ (Int$\rm_{rel}$) & $\lambda$ (Int$\rm_{rel}$) \\ 
  & for non-deuterated & for singly-deuterated & for doubly-deuterated & for triply-deuterated & for quadruply-deuterated\\ \hline
  Anthracene & 13.93~$\mu \rm m$ {(0.56)} & 13.76~$\mu \rm m$ {(0.41)} & 13.65~$\mu \rm m$ {(0.34)} & 13.65~$\mu \rm m$ {(0.4)} & 13.71~$\mu \rm m$ {(0.37)} \\  \hline
  Pentacene & 13.84~$\mu \rm m$ {(0.41)} & 13.78~$\mu \rm m$ {(0.27)} & 13.78~$\mu \rm m$ {(0.26)} & 13.78~$\mu \rm m$ {(0.28)} & 13.78~$\mu \rm m$ {(0.28)} \\ \hline
\end{tabular}
\end{table*}
{The $4.4$ and $4.65~\mu \rm m$ features, which are supposedly the standard tool {for the probe of} deuterium-containing PAHs in the ISM{,} are not detected in abundance with significant intensity.} 
This seeks for a supportive diagnostic tool for deuterium-containing PAHs at {the} longer wavelength side. C$-$D$\rm_{oop}$ modes may serve the purpose if any `unique'  C$-$D$\rm_{oop}$ feature can be found out, which does not interfere with known C$-$H or C$-$C$-$C modes, even if it is weak.
In this report, we have considered substitution at the solo, duet, trio and quartet C$-$H sites of select PAHs by single or multiple Ds to understand the spectral features arising from these species. {Besides} producing C$-$D$\rm_{oop}$ mode{s}, {replacing H} by D at the periphery of PAH modifies the adjacency group of principal C$-$H bond. For example: a duet C$-$H site is converted into a solo C$-$H site when a single D substitutes an H in the same ring producing a solo C$-$H and a solo C$-$D bond.

D atoms substituted on the periphery of PAHs increase the number of observed transitions by modifying the molecule’s mass distribution, which alters vibrational mode couplings and lifts near-degenerate frequencies. This leads to new active modes, frequency shifts, and changes in spectral intensities.
Table \ref{tab1}-\ref{tab4} {summarize} the wavelength position and intensity of originally appearing dominant C$-$H$\rm_{oop}$ modes in that particular molecule before and after deuteration. 
While {the} wavelength positions remain nearly {the} same, the intensities appear to be affected on account of deuterium substitutions. Deuterium {exchange} in the C$-$H site lowers the number of C$-$H bonds in that molecule, resulting into a reduction in {the} overall intensity of C$-$H$\rm_{oop}$. Large PAHs (for example: circumcoronene and circumovalene) are, however, insusceptible to this substitution in terms of intensity and wavelength position. This is possibly due to {the} large number of C$-$H bonds in large PAHs, which are not affected by mere deuteration.

Apart from any changes in the wavelength position and intensity of existing principal C$-$H$\rm_{oop}$ vibrational modes, a few minor features may {rise up} usually between 10~$-$15~$\mu \rm m$ after deuterium substitution. 
{They} are of weak to moderate intensities and are caused by the change in the structure {of} {PAHs}. While in small to medium PAH molecules, these features may appear with visible intensity, spectra of large PAH molecules (C $\geq$ 50) are not greatly affected by mere deuteration and these modes, although present{,} are feeble in intensity. D substitution may {convert} a duet C$-$H group naturally  into a solo C$-$H and a solo C$-$D bond. As a result, a solo C$-$H$\rm_{oop}$ feature {near} 11.2~$\mu \rm m$ starts to appear because of the newly occurring solo C$-$H bond along with the already existing duet C$-$H$\rm_{oop}$ modes. An example is singly-deuterated pyrene or singly-deuterated coronene in Figure~\ref{fig2}b and \ref{fig2}f, in which features {near} $\sim$11.2~$\mu \rm m$ appear due to {a} solo C$-$H$\rm_{oop}$ near the oop modes of {the} remaining duet C$-$H bonds at 11.8$-$12~$\mu \rm m$. This feature was not present before deuterium {replacement}. However, {the} feature is less intense compared to what is obtained from {the} PAH originally comprising of multiple solo C$-$H bonds. 
Other C$-$H groups can also be modified into some other forms, say a trio C$-$H site into a duet or a solo C$-$H site and a quartet C$-$H site into a trio, a duet or a solo C$-$H site with increasing D substitution. Accordingly, {set of features} start to appear resulting from these {changes}. 

\textit{Features due to C$-$D$\rm_{oop}$:} This report presents feature{s} due to C$-$D$\rm_{oop}$ bending mode{s} in deuterated PAHs in addition to the changes in the {overall} spectra upon deuteration, which {are} discussed above. However, {these modes} can partially be {merged} with other C$-$H$\rm_{oop}$ {or C$-$C$-$C$\rm_{oop}$ modes} and cannot be counted as the sole effect of deuterium substitution. 
The C$-$H$\rm_{oop}$ modes occur in a fairly distinctive gap of 11.1$-$13.6~$\mu \rm m$ with increasing adjacency of C$-$H bonds. We can classify the C$-$D groups too as solo, duet, trio and quartet C$-$D group{s} depending upon how many neighbouring C$-$D bonds are attached in the same ring. This will allow us to study the wavelength regions of solo, duet, trio and quartet C$-$D groups. Table~\ref{tab5} lists the positions and intensities of C$-$D$\rm_{oop}$ modes for a range of molecules that are classified in terms of increasing adjacency of C$-$D unit{s}. Modes with Int$\rm_{rel}$ $\geq$ 0.04 are only listed {and those below this are considered as too feeble to be seen}. The last column in the Table~\ref{tab5} indicates the location or adjacency type of C$-$H units at which the substitution by single of multiple D takes place to produce solo, duet, trio and quartet C$-$D units. For example: the location could be a quartet C$-$H site, but single deuterium substitution at this site will convert it into one solo C$-$D unit and another trio C$-$H unit. {Assuming the difference in C$-$H and C$-$D reduced mass (approximate ratio is 1.34), wavelengths corresponding to C$-$D$\rm_{oop}$ vibrational modes can be estimated. This approach suggests that the solo C$-$D$\rm_{oop}$ should be at around 15~$\mu \rm m$, duet C$-$D$\rm_{oop}$ at around 17~$\mu \rm m$, trio C$-$D$\rm_{oop}$ at around 18~$\mu \rm m$ and quartet C$-$D$\rm_{oop}$ slightly further. The results in table~\ref{tab5} are {near} these ranges with small deviations. These deviation{s} could be due to the interaction with other existing C$-$H {or C$-$C$-$C} vibrational modes and a simple estimation will not work to accurately calculate the positions of C$-$D$\rm_{oop}$ vibrational modes.}

{The solo C$-$D$\rm_{oop}$ mode appears in the range of 14.4$-$15.7~$\mu \rm m$ for small to medium PAH samples.}
{With increasing adjacency of C$-$D units, the C$-$D$\rm_{oop}$ modes gradually shift to {the} longer wavelength side towards $\sim$17~$\mu \rm m$.} For coronene and perylene, this even goes to $\sim$19~$\mu \rm m$ due to duet C$-$D$\rm_{oop}$ and trio C$-$D$\rm_{oop}$ modes, respectively with moderate intensities. For very large{,} compact PAHs, like circumcoronene and circumovalene {carrying} solo and duet C$-$D units, {the C$-$D$\rm_{oop}$} feature
is overpowered by other modes and is present {only} with feeble intensities. 
{For circumcoronene, a comparatively weak solo C$-$D$\rm_{oop}$ feature is present near $\sim$~18$\mu \rm m$. Solo C$-$D$\rm_{oop}$ vibrational modes with contributions from other modes also exist} at 14.49 \& 14.65 \& 18.39~$\mu \rm m$ etc. 
in singly-deuterated circumcoronene, although these appear with weak intensities.
This occurs to other large PAHs too, in which {the} presence of large number of C$-$H modes overpower{s} the C$-$D modes. This makes it difficult to probe deuterated features in large PAHs with minimal deuteration.} Unlike C$-$H$\rm_{oop}$ modes, which {have} a distinctive  wavelength range depending upon the adjacency type of C$-$H units, the wavelength range of C$-$D$\rm_{oop}$ modes belonging to different adjacency group{s} may not be as distinctive as C$-$H$\rm_{oop}$ modes, since C$-$D$\rm_{oop}$ modes endure partial contribution{s} from other modes. Nonetheless, the wavelengths of C$-$D$\rm_{oop}$ modes do not overlap with the wavelength range attributed to C$-$H$\rm_{oop}$ modes.
With increasing {adjacency} of C$-$D units from solo to quartet {signifying {the} increasing {number} of deuterium from {no. of} D=1 to 4 in the same benzene ring}, the intensity of these modes increase.
While deuterated features at 4.4 and 4.65~$\mu \rm m$ are detected towards limited sources \citep[]{Doney15}{,} indicating that the incorporation of deuterium in PAHs {could be} rare with high dependency on the local conditions of the source, D may be sequestered {only} in
large PAHs, which is not yet observationally tested{.} {F}or such PAHs, {the} intensities of {the features at} 4.4 and 4.65~$\mu \rm m$ are likely to be minimum. Hence, judging the deuteration process in PAHs only through detections of {the} 4.4 and 4.65~$\mu \rm m$ features might not be appropriate.
With growing size, compact{,} symmetric PAHs tend to have more solo and duet C$-$H units compared to other adjacency type{s} of C$-$H units, limiting the deuteration site in the same ring, which in turn would produce less intense deuterium {features} associated vibrational modes. 
It is worth to mention that we consider cases in which D atoms are attached to {an} aromatic C atom, although, astronomical observations also suggest that D could be attached to an aliphatic C atom in a PAH molecule \citep[]{Peeters04, Onaka14, Mridu15}, and in such a scenario, C$-$D$\rm_{oop}$ modes might occur at other wavelengths. In this paper, we are not exploring the characteristic wavelengths from such species.
\begin{table*}
\centering
\caption{{Band position and relative intensity ($\geq$ 0.04) of the C$-$D$\rm_{oop}$ mode belonging to different adjacency group}}
\label{tab5}
\newcommand{\mc}[3]{\multicolumn{#1}{#2}{#3}}
\begin{tabular}[c]{c|c|c|c|c}
\hline
mode & PAH & $\lambda$ ($\mu \rm m$) & Int$\rm_{rel}$ & location of D substitution \\ \hline
solo C$-$D$\rm_{oop}$ & Anthracene & 15.12 & {0.1} & solo C$-$H site \\
 &  & 15.37 & {0.14} & quartet C$-$H site\\
& Pyrene & 15.56 & {0.04} & duet C$-$H site \\
&  & 15.67 & {0.13} & trio C$-$H site \\
& Perylene & 14.37 & {0.08} & trio C$-$H site \\
 & Pentacene  & 15.46 & {0.05} & quartet C$-$H site  \\ 
    & {Circumcoronene} & {18.39} & {0.04} & {solo C$-$H site} \\  
   & \ & {18.05} & {0.04} & {duet C$-$H site} \\  \hline 
Duet C$-$D$\rm_{oop}$ & Anthracene & 14.24 & {0.06} & quartet C$-$H site \\
 &  & 16.02 & {0.2} & quartet C$-$H site \\
& Pyrene & 15.78 & {0.08} & duet C$-$H site \\
 &  & 16.4 & {0.16} & trio C$-$H site\\
 & Perylene & 15.15 & {0.07} & trio C$-$H site\\
&  & 17.27 & {0.04} & trio C$-$H site\\
 & Pentacene & 16.05 & {0.13} & quartet C$-$H site \\
 & Coronene & 19.15 & {0.13} & duet C$-$H site \\ 
 & Circumcoronene & 18.26 & {0.07} & duet C$-$H site \\  \hline
 Trio C$-$D$\rm_{oop}$ & Anthracene & 16.96 & {0.32} & quartet C$-$H site \\
& Pyrene & 16.89 & {0.2} & trio C$-$H site \\
& Perylene & 15.29 & {0.05} & trio C$-$H site \\
&  & 19.14 & {0.1} & trio C$-$H site \\ 
 & Pentacene & 17 & {0.13} & quartet C$-$H site \\ \hline
  Quartet C$-$D$\rm_{oop}$ & Anthracene & 17.03 & {0.39} & quartet C$-$H site \\
  & Pentacene & 17.05 & {0.13} & quartet C$-$H site \\ \hline
\end{tabular}
\end{table*}

Using the 3.3 and 4.4~$\mu \rm m$ bands, which are direct manifestations of aromatic C$-$H and C$-$D stretching in a PAH molecule, \citet[]{Yang20, Yang21} derived a D/H ratio in interstellar PAHs by computing the C$-$H and C$-$D stretching band strength (per bond) ratio and comparing that with observed intensities of {the} 3.3 and 4.4~$\mu \rm m$ bands. The estimated degree of deuteration in PAHs (fraction of peripheral D atoms attached to aromatic C ring in PAH) was {estimated} to be 2.4\%. In a similar manner, {the} band strength (per bond) of C$-$H$\rm_{oop}$ modes and the analog{o}us C$-$D$\rm_{oop}$ modes may also be useful to estimate {the} degree of deuteration {of} PAHs. From the sample molecules reported in this paper, we {calculate} $\frac{\rm band~strength~of~the~C-D\rm_{oop}~(per~bond)}{\rm band~strength~of~the~C-H\rm_{oop}~(per~bond)}$, which is found to vary among the sample species unlike the intrinsic strength ratio of {the} 3.3 and 4.4~$\mu \rm m$ {bands}. The calculated ratio is in the range of 0.15 to 1.21. While, the band strength of C$-$H$\rm_{oop}$ modes (per bond) is in the range of 12$-$17 km mol$^{-1}$, the band strength of C$-$D$\rm_{oop}$ modes (per bond) spans a wide range of 2$-$17 km mol$^{-1}$. 
Measuring the intrinsic band strength of C$-$D$\rm_{oop}$ modes on a per bond basis may not be as accurate as measuring the intrinsic band strength of C$-$D$\rm_{stretching}$ modes, the reason being {that} C$-$D$\rm_{oop}$ modes are often found to {overlap} with other modes and are difficult to isolate. Even if C$-$D$\rm_{oop}$ modes are detected, these would suggest a lower limit of D/H. This is in contrast to the C$-$H/C$-$D stretching modes, which {do not overlap} with other modes. 

\textit{{Astronomical implications:}} In addition to the well established major mid-infrared PAH bands, Spitzer have also revealed secondary weaker bands spanning in the range of 15$-$20~$\mu \rm m$ with its powerful sensitivity towards several sources 
\citep[]{Werner04, Sellgren07, Boersma10, Peeters12}. The features are mainly at 15.8, 16.4, 17.0, 17.4, 17.8, 18.9~$\mu \rm m${,} exhibiting spatial distributions. Studies following the detection suggested several probable independent species as carriers, 
since these features are not related to each other. The 15$-$20~$\mu \rm m$ wavelength in general corresponds to in-plane and out-of-plane bending modes of the carbon skeleton (C$-$C$-$C). \citet[]{Boersma10} found no correlation between the detected features in the 15$-$20~$\mu \rm m$ region and the mid-IR PAH features except for the 16.4~$\mu \rm m$ feature, which showed correlation with the 6.2 and 7.6/7.8~$\mu \rm m$ bands. PAHs consisting `pendent' rings could produce the 16.4~$\mu \rm m$ feature, however, the absence of another feature at 13.5~$\mu \rm m$ feature expected from the same species poses a question upon this assignment. \citet[]{Peeters12} found correlation of the bands at 15.8~$\mu \rm m$ and 15$-$18~$\mu \rm m$ plateau with the 11.2~$\mu \rm m$ PAH band and {the one at} 16.4~$\mu \rm m$ with the 12.7~$\mu \rm m$ PAH band. Large neutral PAHs are proposed as carriers for {the} 15.8~$\mu \rm m$ band and {the} 15$-$18~$\mu \rm m$ plateau, ionized PAHs for the 16.4~$\mu \rm m$ band, both neutral and cationic PAHs for the 17.8~$\mu \rm m$ band, C$_{60}$ for the 18.9~$\mu \rm m$ and the 17.4~$\mu \rm m$ feature \citep[and references therein]{Peeters12}. Several other candidate carriers, for example: Diamonds, fullerenes, nanoparticles of specific minerology are too {proposed} as carriers for some of these bands \citep[]{Sellgren07, Werner04}. Recent JWST observations of the Orion Bar suggests the presence of a very weak feature at 14.21~$\mu \rm m$ \citep[]{Chown24}. It is still challenging to detect faint features at wavelengths longer than $\sim$13~$\mu \rm m$ even with the JWST. 

Our study presents several {features at} wavelengths between 14$-$19~$\mu \rm m$ due to the C$-$D$\rm_{oop}$ modes {of different C$-$D adjacency group exhibiting weak to moderate intensities.} 
This {suggests} that besides the already proposed carriers for the 15$-$20~$\mu \rm m$ previously, deuterium containing PAHs could also be an elusive candidate carrier for these features. 
{A comparable wavelength match is found for our sample molecules (for example: pyrene, perylene) with observations, which makes them viable to be included in the list of candidate carriers and further investigation is worth considering.}
{{Studies} on a wide range of sample varying in size {are} essential to validate this.}  
{I}t is {important} to mention that isomerism also affects the position and intensity by some extent{, although} in our study, we restrict ourselves to limited PAHs and structures as a preliminary study. 
As an initial study on deuterated PAHs, we also restrict ourselves to small PAHs mostly and to include size effect, we consider only two large PAHs, circumcoronene (C$_{54}$H$_{18}$) and circumovalene (C$_{66}$H$_{20}$) in our sample.
C$-$D$\rm_{oop}$ {modes} in these two molecules give rise to a feature near {$\sim$~18$\mu \rm m$} with much weak{er} relative intensity. There exists a plateau component below the emission of the band in the 10$-$15~$\mu \rm m$ \citep[]{Hony01, Peeters04a}, whose origin remains unclear. {M}inor features due to the combination of C$-$H vibrational modes {newly created} by the change in PAHs' geometry may explain some of these plateau component{s}.

A practical approach to confirm any presence of deuterium-containing PAHs towards astronomical environments is that the characteristic features due to C$-$D$\rm_{stretching}$ and C$-$D$\rm_{oop}$ bending should be observed together. However, detection of C$-$D$\rm_{stretching}$ features at 4.4$-$4.65~$\mu \rm m$ \citep[]{Mridu15} is limited by few observations {\citep[]{Onaka14, Doney15, Boersma2023, peeters2024}}. 
On the other hand, C$-$D$\rm_{oop}$ bending features appear in a range rather than at a discrete position across molecules, making it complex to select the particular wavelengths that would help detection precisely. 
An additional weak to moderate $\sim$11.1~$\mu \rm m$ feature is present resulting from {an} oop mode of a converted solo C$-$H unit when deuteration occurs at a duet, trio and quartet C-H sites in PAHs at subsequent steps to produce a solo C$-$H unit.
Its position may slightly differ from the observed position of 11.2~$\mu \rm m$. PAH molecules with predominant trio and quartet C$-$H sites would acquire a solo C$-$H unit only when more than one deuterium would substitute the C$-$H sites. {O}wing to the rare detection of 4.4$-$4.65~$\mu \rm m$, single deuterium substitution is more likely {to occur compared to multiple deuterium substitution}. 
In order to investigate the possibility of deuteration of PAHs and the extent upto which a molecule can be deuterated, simultaneous evidence of these features may be useful. Previous reports suggest the presence of {a} feature at 11.0$-$11.1~$\mu \rm m$ due to a characteristic band from PAH cations. A detail investigation is pivotal to understand if deuterated PAHs could contribute to this. If we assume minimal deuteration (no. of D=1) resulting into a solo C$-$D, there are additional new features (discussed above) that are nearly equal intense as the C-D$\rm_{oop}$ feature{. H}owever, the position is not necessarily discrete across molecules and depends on shape and size. {The features are marked in the figures \ref{fig1}-\ref{fig4}.} 
For circumcoronene and circumovalene, no such significant feature is observed except the feature near 11.0~$\mu \rm m$ in case of single deuteration at the duet C$-$H site. Investigation on the simultaneous presence or absence of these features together with C$-$D$\rm_{oop}$ features are critical to approve or disapprove the presence of deutered PAHs in the observed target. {Note that compared to the C$-$D stretching mode at 4.4 \,$\mu$m, the C$-$D oop mode can be excited in lower-temperature regions and in larger PAHs, which may provide more favourable conditions for deuteration.} 
\section*{Conclusion:}
Deuterium containing PAH molecules show characteristic C$-$D$\rm_{oop}$ vibrational feature{s} 
depending on its adjacent group, size and \% of deuteration.  These features are moderate in strength.
The range for C$-$H$\rm_{oop}$ is 11.2$-$13.6~$\mu \rm m$ and C$-$D$\rm_{oop}$ is
higher in wavelength than these. While it is difficult to assign a fixed feature in terms of position arising due to C$-$D$\rm_{oop}$ 
and so is for C$-$H$\rm_{oop}$
bending, it is a range of wavelengths at which features due to C$-$D$\rm_{oop}$ or C$-$H$\rm_{oop}$
rise. The range for C$-$D$\rm_{oop}$ is 14$-$19~$\mu \rm m$ with the shorter wavelengths being dominated by solo C$-$D$\rm_{oop}$. This range does not overlap with the range dedicated to the analogous C$-$H$\rm_{oop}$ vibrations. 
The intensity of inherent prevailing C$-$H$\rm_{oop}$ modes of a PAH molecule are affected by the presence of deuterium at its peripheral sites upto some extent. We have compared the spectra of non-deuterated and deuterated PAH molecules to identify such indirect effect{s}. While the spectra of non-deuterated PAH molecule is comparatively simpler, substituting one of the peripheral H atoms at any adjacent C$-$H site by D results into a comparatively complex spectra with several new more or less important characteristic modes.
While the total number of atoms are same in both deuterated and non-deuterated PAHs, deuteration can lead to the resolution of degenerate vibrational modes due to changes in the mass distribution. This resolution breaks some previously degenerate modes into distinct frequencies, increasing the number of observed transitions.

Below are a few observations from the study:
\begin{enumerate}
    \item The positions of existing solo, duet, trio or quartet C$-$H$\rm_{oop}$ {do} not change significantly upon deuteration and turn to appear in close neighbouring wavelengths with a few additional weak features. The shift is within the range of 
    C$-$H$\rm_{oop}$ vibrations of the respective adjacency group.
    \item The intensity of solo, duet, trio or quartet C$-$H$\rm_{oop}$ in non-deuterated PAHs
    reduces upon deuteration to give additional features in the neighbouring wavelengths. While the change in intensity is notable in smaller PAHs, it is comparatively less in large PAH. 
    \item A feature at $\sim$~11.0$-$11.2~$\mu \rm m$ may appear in some deuterated-PAHs{,} which originally do not exhibit a solo C$-$H unit. This feature is due to the creation of a solo C$-$H unit upon deuteration of C$-$H site.
    \item A very weak to moderate C$-$D$\rm_{oop}$ feature is present in all deuterium-conaining PAHs, which is comparatively stronger in intensity in small PAHs as compared to large PAHs.  With increasing deuteration, the feature is enhanced and red shifted in some molecules. Some of these wavelengths lie in the same region as a few detected features in the 15$-$20~$\mu \rm m$ region by Spitzer, JWST etc \citep[]{Peeters12, Chown24}.
    {Further studies are needed to make precise matching.}
    \end{enumerate}

This study adds to our existing knowledge on {the AIB} carries. The region 15$-$20~$\mu \rm m$, which is usually attributed to C$-$C$-$C bending modes may also possess features from C$-$D$\rm_{oop}$ vibrational modes from deuterated PAHs. Also, the well established bands, for example: 11.2~$\mu \rm m$ may also have a contribution from C$-$H$\rm_{oop}$ in a deuterated PAHs, resulting from a modified solo C$-$H unit, if such molecules are present. The observational evidence of features at 14$-$19~$\mu \rm m$ and also at 4.4$-$4.65~$\mu \rm m$, although few, captivates us to include deuterated PAHs in the list of candidate carriers. It is important to mention that deuteration of interstellar PAHs is considered as a rare process and therefore, features from deuterium are also expected to be less in strength. This makes high resolution high spatially resolved spectra a requisite to detect deuterated features, both C$-$D$\rm_{oop}$ and C$-$D$\rm_{stretching}$ in observations. JWST NIRSpec and MIRI spectra may potentially be used to search for these features, if present any.{The favorable low temperature environment for deuteration process to occur efficiently will require the community to look for deuterated features in less excited regions, such as Horsehead nebula \citep[]{Abergel2024}. The Orion Bar may not be a good target to look for C$-$D features, since it may be too warm.} 
{This report is based on the study conducted on a few select PAHs. Theoretical calculations on a wide range of PAHs is necessary for further investigation and explicit conclusion.} 
\section*{Acknowledgements}
MB (Mridusmita Buragohain) thanks the DST INSPIRE Faculty fellowship for awarding her research fellowship and grant. AP (Amit Pathak) acknowledges financial support from the IoE grant of Banaras Hindu University (R/Dev/D/IoE/Incentive/2021-22/32439), financial support through the Core Research Grant of SERB, New Delhi (CRG/2021/000907), financial support from the DST-JSPS grant 2024. MB and AP thank the Inter-University Centre for Astronomy and Astrophysics, Pune for visiting associateship. {TO (Takashi Onaka) acknowledges the support by the Japan Society for the Promotion of Science (JSPS) KAKENHI Grant Number JP24K07087. AV acknowledges an appointment to the NASA Postdoctoral Program at NASA Ames Research Center, administered by the Oak Ridge Associated Universities through a contract with NASA.} 
\section*{DATA AVAILABILITY}
The data underlying this article are available from the authors upon request.
\def\aj{AJ}%
\def\actaa{Acta Astron.}%
\def\araa{ARA\&A}%
\def\apj{ApJ}%
\def\apjl{ApJ}%
\def\apjs{ApJS}%
\def\ao{Appl.~Opt.}%
\def\apss{Ap\&SS}%
\def\aap{A\&A}%
\def\aapr{A\&A~Rev.}%
\def\aaps{A\&AS}%
\def\azh{AZh}%
\def\baas{BAAS}%
\def\bac{Bull. astr. Inst. Czechosl.}%
\def\caa{Chinese Astron. Astrophys.}%
\def\cjaa{Chinese J. Astron. Astrophys.}%
\def\icarus{Icarus}%
\def\jcap{J. Cosmology Astropart. Phys.}%
\def\jrasc{JRASC}%
\def\mnras{MNRAS}%
\def\memras{MmRAS}%
\def\na{New A}%
\def\nar{New A Rev.}%
\def\pasa{PASA}%
\def\pra{Phys.~Rev.~A}%
\def\prb{Phys.~Rev.~B}%
\def\prc{Phys.~Rev.~C}%
\def\prd{Phys.~Rev.~D}%
\def\pre{Phys.~Rev.~E}%
\def\prl{Phys.~Rev.~Lett.}%
\def\pasp{PASP}%
\def\pasj{PASJ}%
\def\qjras{QJRAS}%
\def\rmxaa{Rev. Mexicana Astron. Astrofis.}%
\def\skytel{S\&T}%
\def\solphys{Sol.~Phys.}%
\def\sovast{Soviet~Ast.}%
\def\ssr{Space~Sci.~Rev.}%
\def\zap{ZAp}%
\def\nat{Nature}%
\def\iaucirc{IAU~Circ.}%
\def\aplett{Astrophys.~Lett.}%
\def\apspr{Astrophys.~Space~Phys.~Res.}%
\def\bain{Bull.~Astron.~Inst.~Netherlands}%
\def\fcp{Fund.~Cosmic~Phys.}%
\def\gca{Geochim.~Cosmochim.~Acta}%
\def\grl{Geophys.~Res.~Lett.}%
\def\jcp{J.~Chem.~Phys.}%
\def\jpca{J.~Phys.~Chem.~A}%
\def\jgr{J.~Geophys.~Res.}%
\def\jqsrt{J.~Quant.~Spec.~Radiat.~Transf.}%
\def\memsai{Mem.~Soc.~Astron.~Italiana}%
\def\nphysa{Nucl.~Phys.~A}%
\def\physrep{Phys.~Rep.}%
\def\physscr{Phys.~Scr}%
\def\planss{Planet.~Space~Sci.}%
\def\procspie{Proc.~SPIE}%
\let\astap=\aap
\let\apjlett=\apjl
\let\apjsupp=\apjs
\let\applopt=\ao



\bibliographystyle{mnras}
\bibliography{mridu} 


\appendix
\section{comparison spectra of naphthalene, anthracene and pyrene attained with different methods}
A comparison has been made between gas phase experimental spectra (NIST)  and theoretical spectra of PAHs attained with different methods. These are shown in figures~\ref{fig:fig1}, \ref{fig:fig2}, \ref{fig:fig3}. 
\begin{minipage}{0.5\textwidth}
\includegraphics[scale=0.27]{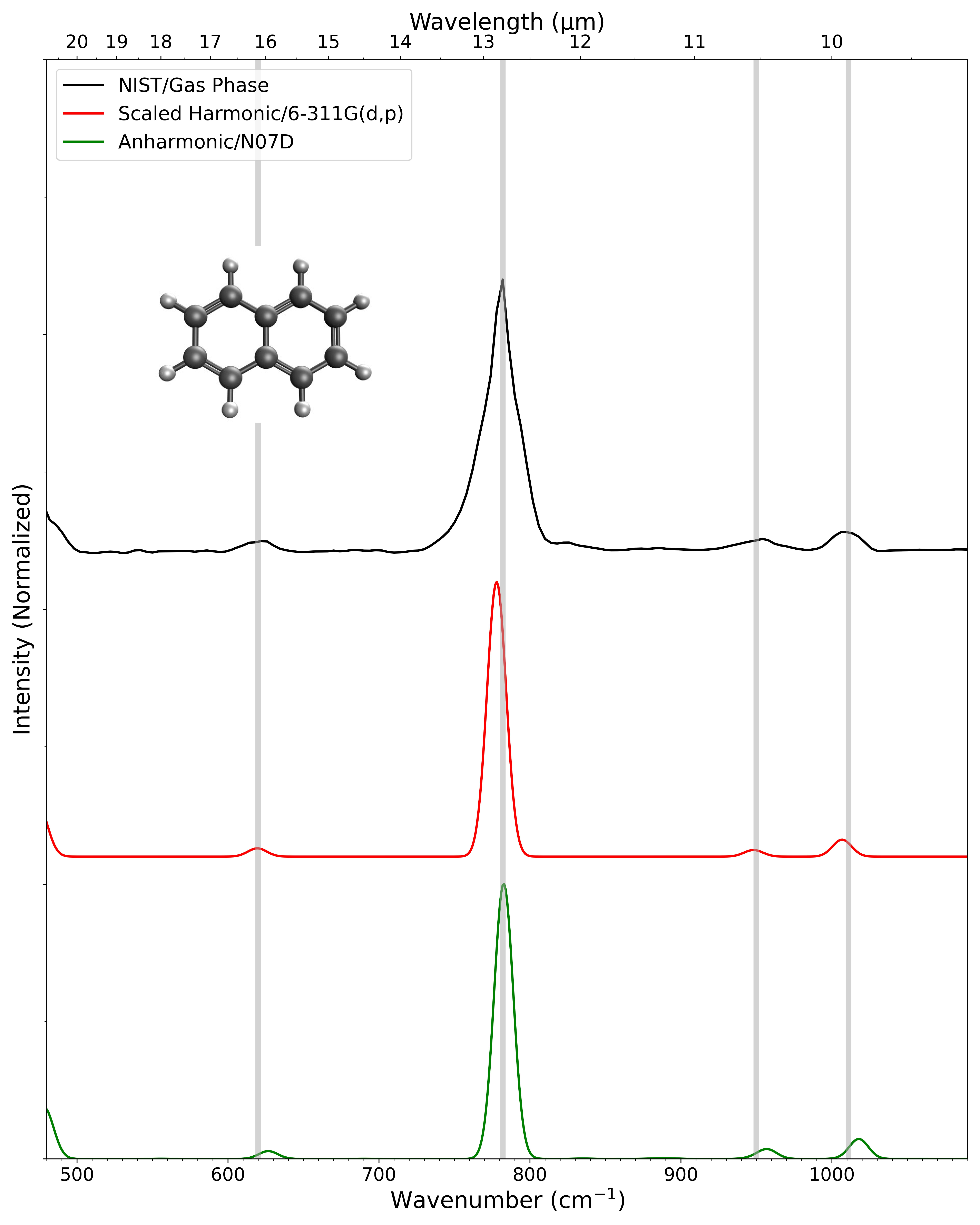}
\captionof{figure}{IR spectra of naphthalene, computed using anharmonic (green) and harmonic methods (red), compared with NIST gas phase spectra (black) at 300 K \citep{NISTIRData2015}. The gray vertical lines are experimental peak positions. The anharmonic and scaled harmonic peak positions are reproduced within 0.8 and 0.5\% of the experiment, respectively.}
\label{fig:fig1} 
\end{minipage}
\begin{minipage}{0.5\textwidth}
\includegraphics[scale=0.27]{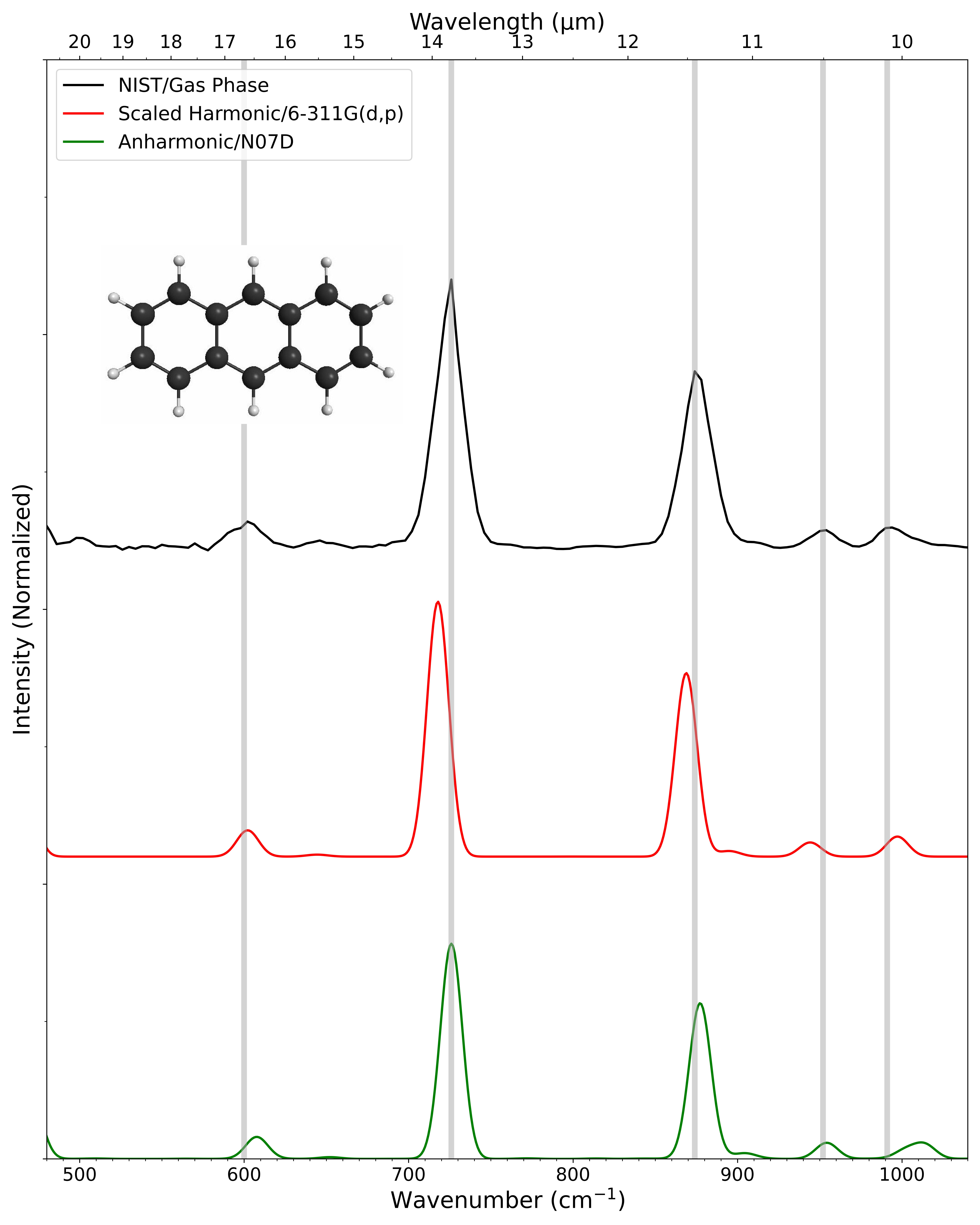}
\captionof{figure}{{IR spectra of anthracene, computed using anharmonic (green) and harmonic methods (red), compared with NIST gas phase spectra (black) at 300 K \citep{NISTIRData2015}. The gray vertical lines are experimental peak positions. The anharmonic and scaled harmonic peak positions are reproduced within 0.9 and 0.7\% of the experiment, respectively.}}
\label{fig:fig2} 
\end{minipage}
\begin{minipage}{0.5\textwidth}
\includegraphics[scale=0.27]{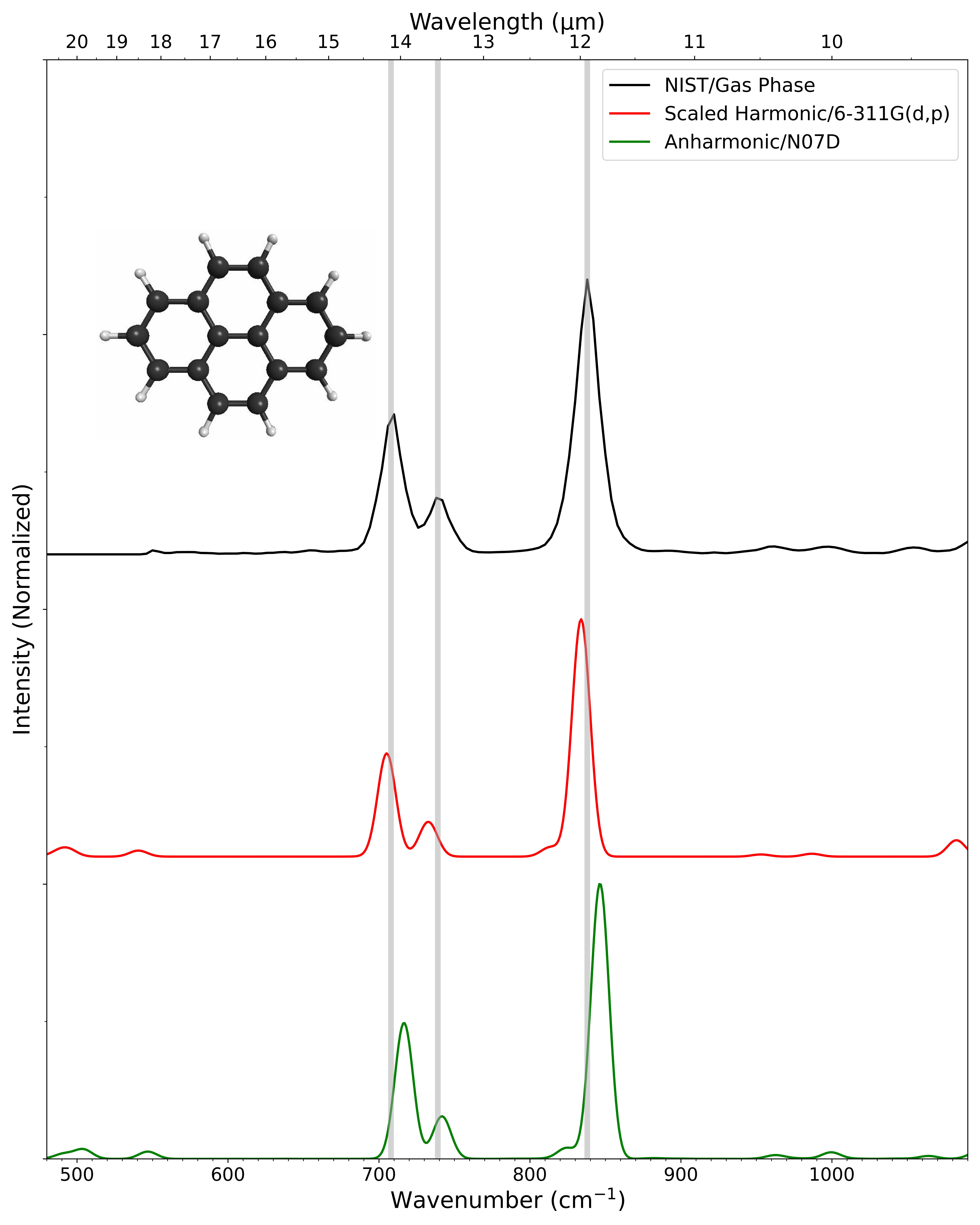}
\captionof{figure}{{IR spectra of pyrene, computed using anharmonic (green) and harmonic methods (red), compared with NIST gas phase spectra (black) at 300 K \citep{NISTIRData2015}. The gray vertical lines are experimental peak positions.
The anharmonic and scaled harmonic peak positions are reproduced within 1.2 and 0.8\% of the experiment, respectively.}}
\label{fig:fig3} 
\end{minipage}
\section{Additional tables}
Below are the tables (\ref{tab:tab1}-\ref{tab:tab33}) for scaled harmonic frequency (cm$^{-1}$) and absolute IR intensity ($\geq$ 0.5 km/mol)  of the sample molecules. 

\begin{minipage}{0.45\textwidth}
\centering
\captionof{table}{{mode number, position and intensity of the vibrational modes in the 1000$-$500 cm$^{-1}$ range in anthracene.}}
\label{tab:tab1}
\begin{tabular}[c]{c|c|c} \hline
mode & frequency & IR intensity \\
 & cm$^{-1}$ & km/mol \\
\hline
$\nu_{34}$ & 997.22 & 6.34 \\ 
$\nu_{37}$ & 944.19	& 4.48 \\
$\nu_{40}$ & 894.91 & 1.77 \\
$\nu_{42}$ & 868.83 & 58.04 \\
$\nu_{50}$ & 717.85	& 80.61 \\
$\nu_{51}$ & 644.45	& 0.63 \\
$\nu_{53}$ & 602.16 & 8.31 \\ \hline
\\
\end{tabular}
\end{minipage}
\begin{minipage}{0.45\textwidth}
\centering
\captionof{table}{{mode number, position and intensity of the vibrational modes in the 1000$-$500 cm$^{-1}$ range in singly-deuterated anthracene. The location of  the substituting D is at solo C$-$H site.}}
\label{tab:tab2}
\begin{tabular}[c]{c|c|c} \hline
mode & frequency & IR intensity \\
 & cm$^{-1}$ & km/mol \\
\hline
$\nu_{37}$ & 942.66	& 2.38 \\
$\nu_{39}$ & 890.43 & 1.76 \\
$\nu_{41}$ & 878.70	& 18.43 \\
$\nu_{43}$ & 835.54	& 22.04 \\
$\nu_{44}$ & 782.25 & 15.58 \\
$\nu_{49}$ & 731.27 & 66.47 \\
$\nu_{50}$ & 661.32	& 13.54 \\
$\nu_{51}$ & 642.91 & 0.60 \\
$\nu_{53}$ & 600.52	& 8.37 \\
$\nu_{54}$ & 568.81 & 0.67 \\
 \hline \\
\end{tabular}
\end{minipage}
\begin{minipage}{0.45\textwidth}
\centering
\captionof{table}{{mode number, position and intensity of the vibrational modes in the 1000$-$500 cm$^{-1}$ range in pentacene.}}
\label{tab:tab3}
\begin{tabular}[c]{c|c|c} \hline
mode & frequency & IR intensity \\
 & cm$^{-1}$ & km/mol \\
\hline
$\nu_{50}$ & 987.84 & 8.09 \\
$\nu_{53}$ & 944.74	& 4.75 \\
$\nu_{56}$ & 898.15	& 2.30 \\
$\nu_{57}$ & 894.17 &	88.15 \\
$\nu_{66}$ & 815.23	& 17.66 \\
$\nu_{73}$ & 725.08	& 3.67 \\
$\nu_{74}$ & 722.78	& 76.07 \\
$\nu_{78}$ & 623.25	& 5.29 \\
$\nu_{80}$ & 564.24	& 2.91 \\ \hline \\
\end{tabular}
\end{minipage}
\begin{minipage}{0.45\textwidth}
\centering
\captionof{table}{{mode number, position and intensity of the vibrational modes in the 1000$-$500 cm$^{-1}$ range in singly-deuterated pentacene. The location of the substituting D is at solo C$-$H site.}}
\label{tab:tab4}
\begin{tabular}[c]{c|c|c} \hline
mode & frequency & IR intensity \\
 & cm$^{-1}$ & km/mol \\
\hline
$\nu_{49}$ & 987.96	& 8.61 \\
$\nu_{53}$ & 944.59	& 3.77 \\ 
$\nu_{56}$ & 898.09 & 2.28 \\
$\nu_{57}$ & 891.50	& 49.35 \\
$\nu_{58}$ & 874.90	& 30.32 \\
$\nu_{59}$ & 874.40	& 3.10 \\
$\nu_{61}$ & 835.26	& 11.30 \\
$\nu_{63}$ & 815.42 & 4.16 \\
$\nu_{67}$ & 766.91	& 4.03 \\
$\nu_{69}$ & 743.84 & 2.66 \\
$\nu_{72}$ & 723.91 & 67.59 \\
$\nu_{73}$ & 722.17 & 3.40 \\
$\nu_{76}$ & 655.56	& 3.20 \\
$\nu_{78}$ & 623.07	& 5.41 \\
$\nu_{80}$ & 564.23 & 2.91 \\
\hline \\
\end{tabular}
\end{minipage}
\begin{minipage}{0.45\textwidth}
\centering
\captionof{table}{{mode number, position and intensity of the vibrational modes in the 1000$-$500 cm$^{-1}$ range in pyrene.}}
\label{tab:tab5}
\begin{tabular}[c]{c|c|c} \hline
mode & frequency & IR intensity \\
 & cm$^{-1}$ & km/mol \\
\hline
$\nu_{37}$ & 986.73	& 1.21 \\
$\nu_{40}$ & 952.93 & 0.90 \\
$\nu_{45}$ & 833.90	& 100.76 \\
$\nu_{47}$ & 813.43	& 3.96 \\
$\nu_{51}$ & 732.69	& 14.76 \\
$\nu_{53}$ & 705.15	& 43.83 \\
$\nu_{58}$ & 540.66	& 2.55 \\
\hline \\
\end{tabular}
\end{minipage}
\begin{minipage}{0.45\textwidth}
\centering
\captionof{table}{{mode number, position and intensity of the vibrational modes in the 1000$-$500 cm$^{-1}$ range in singly-deuterated pyrene. The location of the substituting D is at duet C$-$H site.}}
\label{tab:tab6}
\begin{tabular}[c]{c|c|c} \hline
mode & frequency & IR intensity \\
 & cm$^{-1}$ & km/mol \\
\hline
$\nu_{36}$ & 986.73	&	1.22 \\
$\nu_{37}$ & 962.62	&   0.58 \\
$\nu_{39}$ & 952.93	&	0.89 \\
$\nu_{41}$ & 904.21	&	1.94 \\
$\nu_{42}$ & 902.71	&	7.24 \\
$\nu_{43}$ & 883.83	&	5.05 \\
$\nu_{44}$ & 873.34	&	4.73 \\
$\nu_{45}$ & 825.17	&	33.93 \\
$\nu_{46}$ & 819.25	&	37.54 \\
$\nu_{47}$ & 794.80 &	1.03 \\
$\nu_{48}$ & 786.35	&	2.24 \\
$\nu_{49}$ & 760.04	&	2.77 \\
$\nu_{50}$ & 756.33	&	4.40 \\
$\nu_{52}$ & 716.49	&	41.07 \\
$\nu_{54}$ & 680.45	&	8.98 \\
$\nu_{55}$ & 642.49	&	5.72 \\
$\nu_{58}$ & 539.42 &	2.51 \\
 \hline \\
\end{tabular}
\end{minipage}
\begin{minipage}{0.45\textwidth}
\centering
\captionof{table}{{mode number, position and intensity of the vibrational modes in the 1000$-$500 cm$^{-1}$ range in doubly-deuterated pyrene. The location of the substituting Ds are at duet C$-$H site.}}
\label{tab:tab7}
\begin{tabular}[c]{c|c|c} \hline
mode & frequency & IR intensity \\
 & cm$^{-1}$ & km/mol \\
\hline
$\nu_{35}$ & 986.80	& 1.34 \\
$\nu_{39}$ & 952.83	& 0.65 \\
$\nu_{41}$ & 886.54	& 1.52 \\
$\nu_{43}$ & 835.52 & 2.27 \\
$\nu_{45}$ & 822.97 & 79.27 \\
$\nu_{47}$ & 779.11	& 3.07 \\
$\nu_{49}$ & 757.56	& 7.66 \\
$\nu_{52}$ & 715.30 & 44.50 \\
$\nu_{55}$ & 633.81	& 10.91 \\
$\nu_{58}$ & 538.02 &	2.43 \\
 \hline \\
\end{tabular}
\end{minipage}
\begin{minipage}{0.45\textwidth}
\centering
\captionof{table}{{mode number, position and intensity of the vibrational modes in the 1000$-$500 cm$^{-1}$ range in coronene.}}
\label{tab:tab8}
\begin{tabular}[c]{c|c|c} \hline
mode & frequency & IR intensity \\
 & cm$^{-1}$ & km/mol \\
\hline
$\nu_{59}$ & 851.15 & 154.37 \\
$\nu_{67}$ & 763.74 & 6.47 \\
$\nu_{68}$ & 763.46	& 6.50 \\
$\nu_{79}$ & 550.90	& 38.40 \\
$\nu_{81}$ & 539.63 &	0.75 \\
 \hline \\
\end{tabular}
\end{minipage}
\begin{minipage}{0.45\textwidth}
\centering
\captionof{table}{{mode number, position and intensity of the vibrational modes in the 1000$-$500 cm$^{-1}$ range in singly-deuterated coronene. The location of the substituting D is at duet C$-$H site.}}
\label{tab:tab9}
\begin{tabular}[c]{c|c|c} \hline
mode & frequency & IR intensity \\
 & cm$^{-1}$ & km/mol \\
\hline
$\nu_{56}$ &	926.05 &	0.67 \\
$\nu_{57}$ & 893.24	& 18.23 \\
$\nu_{58}$ & 878.43	& 2.14 \\
$\nu_{59}$ & 842.29	& 116.37 \\
$\nu_{61}$ & 811.67	& 7.00 \\
$\nu_{65}$ & 777.59	& 2.00 \\
$\nu_{66}$ & 764.08	& 6.01 \\
$\nu_{68}$ & 757.29	& 6.08 \\
$\nu_{71}$ &	691.35 &	0.64 \\
$\nu_{80}$ & 545.51	& 31.85 \\
$\nu_{81}$ & 537.91	& 1.64 \\
$\nu_{82}$ & 528.14	& 4.95 \\
$\nu_{83}$ & 506.20	& 2.44 \\
\hline \\
\end{tabular}
\end{minipage}
\begin{minipage}{0.45\textwidth}
\centering
\captionof{table}{{mode number, position and intensity of the vibrational modes in the 1000$-$500 cm$^{-1}$ range in doubly-deuterated coronene. The location of the substituting Ds are at duet C$-$H site.}}
\label{tab:tab10}
\begin{tabular}[c]{c|c|c} \hline
mode & frequency & IR intensity \\
 & cm$^{-1}$ & km/mol \\
\hline
$\nu_{57}$ & 842.81	& 126.00 \\
$\nu_{58}$ & 833.10	& 5.48 \\
$\nu_{60}$ & 812.83	& 7.94 \\
$\nu_{64}$ & 778.38 & 2.23 \\
$\nu_{66}$ & 761.04 & 4.35 \\
$\nu_{67}$ & 751.57	& 6.07 \\
$\nu_{77}$ &	613.07	& 0.93 \\
$\nu_{80}$ & 539.96	& 18.94 \\
$\nu_{81}$ &	531.39 & 0.59 \\
$\nu_{82}$ & 522.21 &	20.87 \\ \hline \\
\end{tabular}
\end{minipage}
\begin{minipage}{0.45\textwidth}
\centering
\captionof{table}{{mode number, position and intensity of the vibrational modes in the 1000$-$500 cm$^{-1}$ range in singly-deuterated pyrene. The location of the substituting D is at trio C$-$H site.}}
\label{tab:tab11}
\begin{tabular}[c]{c|c|c} \hline
mode & frequency & IR intensity \\
 & cm$^{-1}$ & km/mol \\
\hline
$\nu_{36}$ & 980.09	& 1.25 \\
$\nu_{43}$ & 878.17	& 3.36 \\
$\nu_{44}$ & 840.33	& 70.14 \\
$\nu_{45}$ & 826.00 & 18.90 \\
$\nu_{46}$ & 819.03	& 7.27 \\
$\nu_{47}$ & 813.10	& 3.74 \\
$\nu_{50}$ & 745.28	& 9.56 \\
$\nu_{52}$ & 711.85 & 13.26 \\
$\nu_{54}$ & 677.43	& 13.99 \\
$\nu_{55}$ & 638.34	& 17.17 \\
$\nu_{57}$ & 556.94	& 1.30 \\
$\nu_{58}$ & 538.14	& 2.51 \\
$\nu_{59}$ &	515.03 &	0.63 \\
\hline \\
\end{tabular}
\end{minipage}
\begin{minipage}{0.45\textwidth}
\centering
\captionof{table}{{mode number, position and intensity of the vibrational modes in the 1000$-$500 cm$^{-1}$ range in doubly-deuterated pyrene. The location of the substituting Ds are at trio C$-$H site.}}
\label{tab:tab12}
\begin{tabular}[c]{c|c|c} \hline
mode & frequency & IR intensity \\
 & cm$^{-1}$ & km/mol \\
\hline
$\nu_{35}$ &	989.69 &	0.58 \\
$\nu_{36}$ &	978.77 &	1.72 \\
$\nu_{40}$ &	911.23 &	1.73 \\
$\nu_{41}$ &	887.83 &	9.00 \\
$\nu_{42}$ &	879.02 &	8.57 \\
$\nu_{43}$ &	835.60 &	1.84 \\
$\nu_{44}$ &	828.12 &	54.43 \\
$\nu_{45}$ &	819.22 &	8.40 \\
$\nu_{46}$ &	811.87 &	3.52 \\
$\nu_{47}$ &	797.16 &	3.57 \\
$\nu_{49}$ &	753.14 &	10.52 \\
$\nu_{50}$ &	744.60 &	5.08 \\
$\nu_{52}$ &	698.68 &	23.04 \\
$\nu_{54}$ &	667.23 &	1.06 \\
$\nu_{55}$ &	609.75 &	18.57 \\
$\nu_{58}$ &	536.36 &	2.53 \\
$\nu_{59}$ &	514.98 &	0.62 \\
 \hline \\
\end{tabular}
\end{minipage}
\begin{minipage}{0.45\textwidth}
\centering
\captionof{table}{{mode number, position and intensity of the vibrational modes in the 1000$-$500 cm$^{-1}$ range in triply-deuterated pyrene. The location of the substituting Ds are at trio C$-$H site.}}
\label{tab:tab13}
\begin{tabular}[c]{c|c|c} \hline
mode & frequency & IR intensity \\
 & cm$^{-1}$ & km/mol \\
\hline 
$\nu_{35}$ &	972.75 &	1.01 \\
$\nu_{41}$ &	850.09 &	5.00 \\
$\nu_{42}$ &	830.86 &	72.51 \\
$\nu_{43}$ &	826.37 &	0.83 \\
$\nu_{44}$ &	824.45 &	0.51 \\
$\nu_{45}$ &	811.37 &	3.06 \\
$\nu_{47}$ &	773.44 &	13.01 \\
$\nu_{48}$ &	768.25 &	0.92 \\
$\nu_{49}$ &	744.60 &	5.20 \\
$\nu_{52}$ &	697.37 &	20.75 \\
$\nu_{55}$ &	592.14 &	19.36 \\
$\nu_{57}$ &	539.05 &	3.97 \\
$\nu_{58}$ &	534.06 &	2.49 \\
 \hline \\
\end{tabular}
\end{minipage}
\begin{minipage}{0.45\textwidth}
\centering
\captionof{table}{{mode number, position and intensity of the vibrational modes in the 1000$-$500 cm$^{-1}$ range in perylene.}}
\label{tab:tab14}
\begin{tabular}[c]{c|c|c} \hline
mode & frequency & IR intensity \\
 & cm$^{-1}$ & km/mol \\
\hline 
$\nu_{47}$ &	958.50 &	0.85 \\
$\nu_{57}$ &	808.28 &	7.31 \\
$\nu_{58}$ &	800.79 &	100.78 \\
$\nu_{60}$ &	784.76 &	7.36 \\
$\nu_{61}$ &	766.25 &	2.83 \\
$\nu_{62}$ &	757.60 &	76.94 \\
$\nu_{70}$ &	575.68 & 3.45 \\
$\nu_{71}$ &	542.45	& 10.28 \\
$\nu_{74}$ &	528.94	& 0.92 \\
\hline \\
\end{tabular}
\end{minipage}
\begin{minipage}{0.45\textwidth}
\centering
\captionof{table}{{mode number, position and intensity of the vibrational modes in the 1000$-$500 cm$^{-1}$ range in singly-deuterated perylene. The location of the substituting D is at trio C$-$H site.}}
\label{tab:tab15}
\begin{tabular}[c]{c|c|c} \hline
mode & frequency & IR intensity \\
 & cm$^{-1}$ & km/mol \\
\hline 
$\nu_{49}$ &	942.50 &	0.53 \\
$\nu_{51}$ &	906.55 &	4.96 \\
$\nu_{55}$ &	828.56 &	8.97 \\
$\nu_{56}$ &	818.28 &	49.53 \\
$\nu_{57}$ &	803.71 &	3.60 \\
$\nu_{58}$ &	786.99 &	23.03 \\
$\nu_{59}$ &	786.89 &	2.66 \\
$\nu_{60}$ &	769.38 &	4.71 \\
$\nu_{61}$ &	764.53 &	4.24 \\
$\nu_{62}$ &	757.01 &	22.20 \\
$\nu_{63}$ &	753.25 &	50.19 \\
$\nu_{65}$ &	695.82 &	13.58 \\
$\nu_{69}$ &	584.08 &	2.79 \\
$\nu_{70}$ &	574.08 &	3.32 \\
$\nu_{71}$ &	542.43 &	10.48 \\
$\nu_{74}$ &	525.94 &	0.60 \\
\hline \\
\end{tabular}
\end{minipage}
\begin{minipage}{0.45\textwidth}
\centering
\captionof{table}{{mode number, position and intensity of the vibrational modes in the 1000$-$500 cm$^{-1}$ range in doubly-deuterated perylene. The location of the substituting Ds are at trio C$-$H site.}}
\label{tab:tab16}
\begin{tabular}[c]{c|c|c} \hline
mode & frequency & IR intensity \\
 & cm$^{-1}$ & km/mol \\
\hline 
$\nu_{44}$ &	993.00 &	1.52 \\ 
$\nu_{45}$ &	956.78 &	0.58 \\
$\nu_{47}$ &	950.33 &	2.31 \\
$\nu_{49}$ &	923.66 &	2.50 \\
$\nu_{50}$ &	893.29 &	5.07 \\
$\nu_{51}$ &	887.37 &	1.46 \\
$\nu_{52}$ &	873.21 &	5.01 \\
$\nu_{53}$ &	865.13 &	1.70 \\
$\nu_{54}$ &	855.48 &	0.82 \\
$\nu_{55}$ &	822.05 &	13.09 \\
$\nu_{56}$ &	797.24 &	5.66 \\
$\nu_{57}$ &	789.75 &	45.10 \\
$\nu_{58}$ &	786.62 &	2.01 \\
$\nu_{59}$ &	774.10 &	2.81 \\
$\nu_{60}$ &	761.60 &	4.01 \\
$\nu_{61}$ &	755.85 &	3.65 \\
$\nu_{62}$ &	754.16 &	69.71 \\
$\nu_{64}$ &	730.95 &	0.70 \\
$\nu_{65}$ &	660.27 &	10.63 \\
$\nu_{68}$ &	612.74 &	1.65 \\
$\nu_{69}$ &	579.15 &	5.25 \\
$\nu_{70}$ &	571.20 &	3.27 \\
$\nu_{71}$ &	542.43 &	10.45 \\
$\nu_{74}$ &	523.40 &	0.54 \\
\hline \\
\end{tabular}
\end{minipage}
\begin{minipage}{0.45\textwidth}
\centering
\captionof{table}{{mode number, position and intensity of the vibrational modes in the 1000$-$500 cm$^{-1}$ range in triply-deuterated perylene. The location of the substituting Ds are at trio C$-$H site.}}
\label{tab:tab17}
\begin{tabular}[c]{c|c|c} \hline
mode & frequency & IR intensity \\
 & cm$^{-1}$ & km/mol \\
\hline 
$\nu_{48}$ &	920.26 &	0.67 \\ 
$\nu_{51}$ &	867.34 &	6.64 \\
$\nu_{53}$ &	840.10 &	1.39 \\
$\nu_{54}$ &	825.77 &	2.08 \\
$\nu_{55}$ &	801.92 &	53.80 \\
$\nu_{56}$ &	796.71 &	5.70 \\
$\nu_{57}$ &	786.23 &	2.55 \\
$\nu_{58}$ &	776.24 &	10.42 \\
$\nu_{59}$ &	758.96 &	2.43 \\
$\nu_{60}$ &	754.29 &	69.18 \\
$\nu_{61}$ &	752.74 &	1.84 \\
$\nu_{62}$ &	748.48 &	3.72 \\
$\nu_{64}$ &	714.72 &	3.03 \\
$\nu_{65}$ &	654.11 &	6.53 \\
$\nu_{68}$ &	611.60 &	0.86 \\
$\nu_{69}$ &    569.17 &	3.07 \\
$\nu_{70}$ &	565.91 &	0.83 \\
$\nu_{73}$ &	522.56 &	14.01 \\
$\nu_{74}$ &	521.91 &	0.58 \\
$\nu_{75}$ &	506.10 &	2.22 \\
\hline \\
\end{tabular}
\end{minipage}
\begin{minipage}{0.45\textwidth}
\centering
\captionof{table}{{mode number, position and intensity of the vibrational modes in the 1000$-$500 cm$^{-1}$ range in singly-deuterated anthracene. The location of the substituting D is at quartet C$-$H site.}}
\label{tab:tab18}
\begin{tabular}[c]{c|c|c} \hline
mode & frequency & IR intensity \\
 & cm$^{-1}$ & km/mol \\
\hline 
$\nu_{33}$ &	999.69 &	3.26 \\
$\nu_{35}$ &	959.66 &	0.98 \\
$\nu_{36}$ &	942.05 &	2.12 \\
$\nu_{37}$ &	905.37 &	0.76 \\
$\nu_{39}$ &	900.62 &	2.94 \\
$\nu_{41}$ &	868.71 &	58.50 \\
$\nu_{42}$ &	859.41 &	1.59 \\
$\nu_{44}$ &	791.60 &	4.65 \\
$\nu_{46}$ &	755.61 &	1.17 \\
$\nu_{47}$ &	742.15 &	2.00 \\
$\nu_{49}$ &	726.70 &	48.16 \\
$\nu_{50}$ &	650.79 &	15.94 \\
$\nu_{51}$ &	643.91 &	0.65 \\
$\nu_{53}$ &	598.79 &	8.01 \\
 \hline \\
\end{tabular}
\end{minipage}
\begin{minipage}{0.45\textwidth}
\centering
\captionof{table}{{mode number, position and intensity of the vibrational modes in the 1000$-$500 cm$^{-1}$ range in doubly-deuterated anthracene. The location of the substituting Ds are at quartet C$-$H site.}}
\label{tab:tab19}
\begin{tabular}[c]{c|c|c} \hline
mode & frequency & IR intensity \\
 & cm$^{-1}$ & km/mol \\
\hline 
$\nu_{32}$ &	999.78 &	3.09 \\
$\nu_{35}$ &	941.96 &	2.20 \\
$\nu_{36}$ &	906.91 &	0.97 \\
$\nu_{37}$ &	902.94 &	2.46 \\
$\nu_{38}$ &	882.54 &	0.63 \\
$\nu_{39}$ &	872.61 &	54.90 \\
$\nu_{40}$ &	872.41 &	2.45 \\
$\nu_{41}$ &	831.80 &	1.19 \\
$\nu_{43}$ &	806.79 &	0.61 \\
$\nu_{44}$ &	786.06 &	1.79 \\
$\nu_{48}$ &	732.44 &	35.42 \\
$\nu_{49}$ &	702.31 &	6.43 \\
$\nu_{50}$ &	637.25 &	0.51 \\
$\nu_{51}$ &	624.18 & 20.35 \\
$\nu_{53}$ &	596.20  & 7.98 \\
$\nu_{54}$ &	558.39	& 1.12 \\
\hline \\
\end{tabular}
\end{minipage}
\begin{minipage}{0.45\textwidth}
\centering
\captionof{table}{{mode number, position and intensity of the vibrational modes in the 1000$-$500 cm$^{-1}$ range in triply-deuterated anthracene. The location of the substituting Ds are at quartet C$-$H site.}}
\label{tab:tab20}
\begin{tabular}[c]{c|c|c} \hline
mode & frequency & IR intensity \\
 & cm$^{-1}$ & km/mol \\
\hline 
$\nu_{31}$ &	999.77 &	3.10 \\ 
$\nu_{33}$ &	944.15 &	1.12 \\
$\nu_{34}$ &	942.20 &	2.73 \\
$\nu_{35}$ &	904.13 &	18.50 \\
$\nu_{36}$ &	894.23 &	0.57 \\
$\nu_{37}$ &	883.23 &	3.31 \\
$\nu_{38}$ &	879.87 &	9.67 \\
$\nu_{39}$ &	860.56 &	22.36 \\
$\nu_{40}$ &	840.09 &	2.40 \\
$\nu_{47}$ &	732.38 &	36.16 \\
$\nu_{49}$ &	682.07 &	1.01 \\
$\nu_{52}$ &	593.38 &	8.00 \\
$\nu_{53}$ &	589.48 &	23.46 \\
\hline \\
\end{tabular}
\end{minipage}
\begin{minipage}{0.45\textwidth}
\centering
\captionof{table}{{mode number, position and intensity of the vibrational modes in the 1000$-$500 cm$^{-1}$ range in quadruply-deuterated anthracene. The location of the substituting Ds are at quartet C$-$H site.}}
\label{tab:tab21}
\begin{tabular}[c]{c|c|c} \hline
mode & frequency & IR intensity \\
 & cm$^{-1}$ & km/mol \\
\hline 
$\nu_{30}$ &	999.77 & 3.10 \\ 
$\nu_{33}$ &	942.09 &	2.16 \\
$\nu_{36}$ &	879.00 &	3.89 \\
$\nu_{37}$ &	870.83 &	45.92 \\
$\nu_{38}$ &	843.78 &	3.40 \\
$\nu_{41}$ &	826.42 &	1.32 \\
$\nu_{44}$ &	758.12 &	6.82 \\
$\nu_{47}$ &	729.66 &	28.64 \\
$\nu_{51}$ &	619.61 &	0.77 \\
$\nu_{52}$ &	589.03 &	7.49 \\
$\nu_{53}$ &	587.09 &	22.93 \\
\hline \\
\end{tabular}
\end{minipage}
\begin{minipage}{0.45\textwidth}
\centering
\captionof{table}{{mode number, position and intensity of the vibrational modes in the 1000$-$500 cm$^{-1}$ range in singly-deuterated pentacene. The location of the substituting D is at quartet C$-$H site.}}
\label{tab:tab22}
\begin{tabular}[c]{c|c|c} \hline
mode & frequency & IR intensity \\
 & cm$^{-1}$ & km/mol \\
\hline 
$\nu_{49}$ &	988.96 &	4.25 \\ 
$\nu_{51}$ &	959.53 &	0.88 \\
$\nu_{52}$ &	944.61 &	2.38 \\
$\nu_{53}$ &	912.22 &	0.68 \\
$\nu_{54}$ &	904.88 &	4.22 \\
$\nu_{55}$ &	899.84 &	3.48 \\
$\nu_{56}$ &	893.48 &	84.70 \\
$\nu_{57}$ &	887.88 &	0.92 \\
$\nu_{59}$ &	871.82 &	1.37 \\
$\nu_{64}$ &	815.56 &	17.22 \\
$\nu_{66}$ &	786.19 &	4.16 \\
$\nu_{69}$ &	744.16 &	5.08 \\
$\nu_{71}$ &	730.90 &	0.94 \\
$\nu_{72}$ &	725.68 &	44.84 \\
$\nu_{73}$ &	718.19 &	3.76 \\
$\nu_{75}$ &	698.10 &	2.92 \\
$\nu_{76}$ &	646.61 &	8.65 \\
$\nu_{78}$ &	620.34 &	5.28 \\
$\nu_{80}$ &	563.99 &	2.89 \\
\hline \\
\end{tabular}
\end{minipage}
\begin{minipage}{0.45\textwidth}
\centering
\captionof{table}{{mode number, position and intensity of the vibrational modes in the 1000$-$500 cm$^{-1}$ range in doubly-deuterated pentacene. The location of the substituting Ds are at quartet C$-$H site.}}
\label{tab:tab23}
\begin{tabular}[c]{c|c|c} \hline
mode & frequency & IR intensity \\
 & cm$^{-1}$ & km/mol \\
\hline 
$\nu_{48}$ &	989.26 &	4.07 \\
$\nu_{51}$ &	944.58 &	2.43 \\
$\nu_{52}$ &	911.78 &	0.74 \\
$\nu_{53}$ &	899.30 &	3.75 \\
$\nu_{54}$ &	895.16 &	88.32 \\
$\nu_{56}$ &	878.27 &	2.26 \\
$\nu_{61}$ &	830.58 &	0.94 \\
$\nu_{62}$ &	820.88 &	14.52 \\
$\nu_{63}$ &	818.17 &	0.51 \\
$\nu_{66}$ &	782.32 &	2.00 \\
$\nu_{71}$ &	725.68 &	38.02 \\
$\nu_{72}$ &	717.84 &	3.63 \\
$\nu_{73}$ &	712.95 &	2.87 \\
$\nu_{75}$ &	682.59 &	3.11 \\
$\nu_{77}$ &	623.01 &	14.94 \\
$\nu_{78}$ &	619.21 &	5.29 \\
$\nu_{80}$ &	560.97 &	2.76 \\
\hline \\
\end{tabular}
\end{minipage}
\begin{minipage}{0.45\textwidth}
\centering
\captionof{table}{{mode number, position and intensity of the vibrational modes in the 1000$-$500 cm$^{-1}$ range in triply-deuterated pentacene. The location of the substituting Ds are at quartet C$-$H site.}}
\label{tab:tab24}
\begin{tabular}[c]{c|c|c} \hline
mode & frequency & IR intensity \\
 & cm$^{-1}$ & km/mol \\
\hline 
$\nu_{47}$ &	989.26 &	4.09 \\
$\nu_{49}$ &	944.58 &	2.48 \\
$\nu_{50}$ &	940.25 &	2.86 \\
$\nu_{51}$ &	906.17 &	44.47 \\
$\nu_{53}$ &	891.54 & 38.87 \\
$\nu_{54}$ &	890.59 &	3.65 \\
$\nu_{55}$ &	886.22 &	5.01 \\
$\nu_{57}$ &	864.64 &	0.63 \\
$\nu_{58}$ &	857.68 &	1.08 \\
$\nu_{60}$ &	832.56 &	1.57 \\
$\nu_{63}$ &	813.07 &	7.38 \\
$\nu_{71}$ &	725.68 &	38.25 \\
$\nu_{72}$ &	713.79 &	3.60 \\
$\nu_{75}$ &	675.11 &	1.74 \\
$\nu_{77}$ &	617.81 &	5.31 \\
$\nu_{79}$ &	588.34 &	18.26 \\
$\nu_{80}$ &	558.07 &	2.66 \\
\hline \\
\end{tabular}
\end{minipage}
\begin{minipage}{0.45\textwidth}
\centering
\captionof{table}{{mode number, position and intensity of the vibrational modes in the 1000$-$500 cm$^{-1}$ range in quadruply-deuterated pentacene. The location of the substituting Ds are at quartet C$-$H site.}}
\label{tab:tab25}
\begin{tabular}[c]{c|c|c} \hline
mode & frequency & IR intensity \\
 & cm$^{-1}$ & km/mol \\
\hline 
$\nu_{47}$ &	989.26 &	4.09 \\
$\nu_{49}$ &	944.58 &	2.43 \\
$\nu_{50}$ &	905.21 &	0.60 \\
$\nu_{51}$ &	895.06 &	82.64 \\
$\nu_{52}$ &	890.51 &	4.95 \\
$\nu_{55}$ &	866.47 &	1.26 \\
$\nu_{56}$ &	862.67 &	0.89 \\
$\nu_{58}$ &	830.99 &	1.63 \\
$\nu_{59}$ &	830.34 &	1.27 \\
$\nu_{62}$ &	815.06 &	9.00 \\
$\nu_{66}$ &	752.47 &	1.39 \\
$\nu_{71}$ &	725.58 &	37.17 \\
$\nu_{72}$ &	705.70 &	3.62 \\
$\nu_{77}$ &	615.85 &	5.14 \\
$\nu_{78}$ &	589.50 &	0.64 \\
$\nu_{79}$ &	586.42 &	17.70 \\
$\nu_{80}$ &	557.87 &	2.62 \\
\hline \\
\end{tabular}
\end{minipage}
\begin{minipage}{0.45\textwidth}
\centering
\captionof{table}{{mode number, position and intensity of the vibrational modes in the 1000$-$500 cm$^{-1}$ range in  circumcoronene.}}
\label{tab:tab26}
\begin{tabular}[c]{c|c|c} \hline
mode & frequency & IR intensity \\
 & cm$^{-1}$ & km/mol \\
\hline 
$\nu_{98}$ &	981.73 &	0.89 \\
$\nu_{99}$ & 981.65 &	0.95 \\
$\nu_{108}$ &	882.49 &	177.33 \\
$\nu_{116}$ &	829.58 &	5.13 \\
$\nu_{117}$ &	829.52 &	5.25 \\
$\nu_{128}$ &	774.13 &	36.06 \\
$\nu_{138}$ &	717.59 &	28.29 \\
$\nu_{146}$ &	642.50 &	4.47 \\
$\nu_{147}$ &	642.46 &	4.54 \\
$\nu_{160}$ &	564.27 &	34.46 \\
\hline \\
\end{tabular}
\end{minipage}
\begin{minipage}{0.45\textwidth}
\centering
\captionof{table}{{mode number, position and intensity of the vibrational modes in the 1000$-$500 cm$^{-1}$ range in  singly-deuterated circumcoronene.  The location of the substituting D is at solo C-H site.}}
\label{tab:tab27}
\begin{tabular}[c]{c|c|c} \hline
mode & frequency & IR intensity \\
 & cm$^{-1}$ & km/mol \\
\hline 
$\nu_{98}$ & 979.84 &	0.87 \\
$\nu_{99}$ &	978.59 &	0.51 \\
$\nu_{108}$ &	881.45 &	142.57 \\
$\nu_{111}$ &	873.88 &	12.30 \\
$\nu_{113}$ &	864.65 &	1.11 \\
$\nu_{114}$ &	844.79 &	0.83 \\
$\nu_{115}$ &	829.00 &	4.26 \\
$\nu_{117}$ &	824.25 &	0.89 \\
$\nu_{118}$ &	819.73 &	3.29 \\
$\nu_{119}$ &	814.68 &	3.28 \\
$\nu_{120}$ &	813.80 &	11.97 \\
$\nu_{123}$ &	797.09 &	0.56 \\
$\nu_{124}$ &	795.04 &	5.33 \\
$\nu_{125}$ &	784.99 &	2.99 \\
$\nu_{127}$ &	777.45 &	25.75 \\
$\nu_{128}$ &	762.02 &	2.91 \\
$\nu_{129}$ &	751.67 &	1.58 \\
$\nu_{137}$ &	718.25 &	28.58 \\
$\nu_{142}$ &	667.54 &	0.71 \\
$\nu_{145}$ &	644.77 &	0.54 \\
$\nu_{146}$ &	640.56 &	4.16 \\
$\nu_{147}$ &	639.61 &	4.13 \\
$\nu_{155}$ &	591.07 &	1.79 \\
$\nu_{160}$ &	563.82 &	22.81 \\
$\nu_{164}$ &	543.64 &	10.68 \\
\hline \\
\end{tabular}
\end{minipage}
\begin{minipage}{0.45\textwidth}
\centering
\captionof{table}{{mode number, position and intensity of the vibrational modes in the 1000$-$500 cm$^{-1}$ range in circumovalene.}}
\label{tab:tab28}
\begin{tabular}[c]{c|c|c} \hline
mode & frequency & IR intensity \\
 & cm$^{-1}$ & km/mol \\
\hline 
$\nu_{119}$ &	975.50 &	0.63 \\
$\nu_{127}$ &	905.85 &	4.39 \\
$\nu_{129}$ &	885.47 &	3.30 \\
$\nu_{130}$ &	885.27 &	179.81 \\
$\nu_{133}$ &	875.32 &	16.31 \\
$\nu_{137}$ &	848.50 &	6.96 \\
$\nu_{144}$ &	805.14 &	1.34 \\
$\nu_{148}$ &	779.87 &	42.19 \\
$\nu_{152}$ &	773.80 &	4.04 \\
$\nu_{157}$ &	741.80 &	0.57 \\
$\nu_{159}$ &	738.16 &	7.04 \\
$\nu_{160}$ &	733.64 &	1.73 \\
$\nu_{163}$ &	719.54 &	1.80 \\
$\nu_{165}$ &	706.80 &	5.81 \\
$\nu_{168}$ &	679.46 & 1.37 \\
$\nu_{170}$ &	677.53 &	6.28 \\
$\nu_{173}$ &	667.18 &	0.82 \\
$\nu_{176}$ &	632.85 &	1.34 \\
$\nu_{178}$ &	624.09 &	3.78 \\
$\nu_{180}$ &	611.12 &	3.75 \\
$\nu_{185}$ &	596.87 &	24.26 \\
$\nu_{193}$ &	563.96 &	11.79 \\
$\nu_{201}$ &	514.04 &	0.69 \\
$\nu_{202}$ &	511.21 &	1.74 \\
\hline \\
\end{tabular}
\end{minipage}
\begin{minipage}{0.45\textwidth}
\centering
\captionof{table}{{mode number, position and intensity of the vibrational modes in the 1000$-$500 cm$^{-1}$ range in  singly-deuterated circumovalene. The location of the substituting D is at solo C-H site.}}
\label{tab:tab29}
\begin{tabular}[c]{c|c|c} \hline
mode & frequency & IR intensity \\
 & cm$^{-1}$ & km/mol \\
\hline 
$\nu_{116}$ &	998.98 &	0.97 \\
$\nu_{118}$ &	973.10 &	0.51 \\
$\nu_{127}$ &	904.70 &	4.94 \\
$\nu_{129}$ &	885.16 &	1.21 \\
$\nu_{130}$ &	884.71 &	118.38 \\
$\nu_{131}$ &	879.57 &	7.34 \\
$\nu_{132}$ &	877.75 &	54.94 \\
$\nu_{135}$ & 868.63 &	1.33 \\
$\nu_{136}$ &	862.96 &	4.45 \\
$\nu_{137}$ &	846.15 &	4.64 \\
$\nu_{139}$ &	830.76 &	0.50 \\
$\nu_{140}$ &	826.65 &	0.81 \\
$\nu_{142}$ &	814.09 &	0.59 \\
$\nu_{143}$ &	809.52 &	5.05 \\
$\nu_{145}$ &	794.91 &	1.19 \\
$\nu_{146}$ &	788.94 &	1.45 \\
$\nu_{147}$ &	780.20 &	36.95 \\
$\nu_{148}$ &	777.09 &	2.80 \\
$\nu_{150}$ & 774.03 &	1.00 \\
$\nu_{151}$ &	773.26 &	2.79 \\
$\nu_{157}$ &	741.69 &	0.54 \\
$\nu_{158}$ &	738.53 &	6.51 \\
$\nu_{164}$ &	707.49 &	3.01 \\
$\nu_{165}$ &	706.12 &	1.84 \\
$\nu_{167}$ &	688.23 &	5.17 \\
$\nu_{168}$ &	678.04 &	1.31 \\
$\nu_{169}$ &	677.87 &	1.10 \\
$\nu_{170}$ &	673.65 &	2.22 \\
$\nu_{176}$ &	631.18 &	1.28 \\
$\nu_{177}$ &	628.03 &	0.53 \\
$\nu_{179}$ &	620.43 &	2.81 \\
$\nu_{180}$ &	610.38 &	3.78 \\
$\nu_{184}$ &	596.91 &	24.51 \\
$\nu_{188}$ &	586.32 &	0.81 \\
$\nu_{194}$ &	557.56 &	2.71 \\
$\nu_{195}$ &	555.12 &	3.93 \\
$\nu_{198}$ &	537.88 &	5.79 \\
$\nu_{202}$ &	509.70 &	1.54 \\
\hline \\
\end{tabular}
\end{minipage}
\begin{minipage}{0.45\textwidth}
\centering
\captionof{table}{{mode number, position and intensity of the vibrational modes in the 1000$-$500 cm$^{-1}$ range in  singly-deuterated circumcoronene. The location of the substituting D is at duet C-H site.}}
\label{tab:tab30}
\begin{tabular}[c]{c|c|c} \hline
mode & frequency & IR intensity \\
 & cm$^{-1}$ & km/mol \\
\hline 
$\nu_{97}$ &	983.32 &	1.93 \\
$\nu_{98}$ &	981.12 &	0.87 \\
$\nu_{106}$ &	905.72 &	20.97 \\
$\nu_{107}$ &	904.20 &	1.98 \\
$\nu_{108}$ &	881.47 &	140.39 \\
$\nu_{110}$ &	874.19 &	9.38 \\
$\nu_{113}$ &	864.82 &	0.99 \\
$\nu_{114}$ &	859.19 &	1.51 \\
$\nu_{116}$ &	829.53 &	5.16 \\
$\nu_{117}$ &	827.68 &	4.35 \\
$\nu_{118}$ &	827.45 &	0.72 \\
$\nu_{124}$ &	791.71 &	1.70 \\
$\nu_{127}$ &	776.53 &	26.97 \\
$\nu_{137}$ &	724.07 &	8.15 \\
$\nu_{139}$ &	704.67 &	18.05 \\
$\nu_{141}$ &	674.55 &	3.25 \\
$\nu_{143}$ &	666.29 &	0.62 \\
$\nu_{145}$ &	644.74 &	0.56 \\
$\nu_{146}$ &	642.42 &	4.52 \\
$\nu_{147}$ &	639.85 &	3.91 \\
$\nu_{158}$ &	577.07 &	0.55 \\
$\nu_{159}$ &	568.25 &	4.08 \\
$\nu_{160}$ &	563.65 &	21.31 \\
$\nu_{164}$ &	554.05 &	9.15 \\
$\nu_{168}$ &	523.44 &	1.03 \\
\hline \\
\end{tabular}
\end{minipage}
\begin{minipage}{0.45\textwidth}
\centering
\captionof{table}{{mode number, position and intensity of the vibrational modes in the 1000$-$500 cm$^{-1}$ range in doubly-deuterated circumcoronene. The location of the substituting Ds are at duet C-H site.}}
\label{tab:tab31}
\begin{tabular}[c]{c|c|c} \hline
mode & frequency & IR intensity \\
 & cm$^{-1}$ & km/mol \\
\hline 
$\nu_{96}$ &	983.98 &	3.66 \\
$\nu_{97}$ &	982.16 &	1.46 \\
$\nu_{105}$ &	943.96 &	1.49 \\
$\nu_{106}$ &	881.64 &	153.89 \\
$\nu_{108}$ &	875.28 &	11.08 \\
$\nu_{111}$ &	866.18 &	1.84 \\
$\nu_{114}$ &	833.62 &	5.54 \\
$\nu_{115}$ &	829.38 &	3.38 \\
$\nu_{117}$ &	826.23 &	5.01 \\
$\nu_{122}$ &	802.07 &	0.70 \\
$\nu_{123}$ &	791.79 &	2.30 \\
$\nu_{125}$ &	784.25 &	0.76 \\
$\nu_{126}$ &	776.54 &	26.99 \\
$\nu_{130}$ & 746.84 &	3.75 \\
$\nu_{134}$ &	744.45 &	1.76 \\
$\nu_{136}$ &	727.29 &	2.14 \\
$\nu_{137}$ &	723.76 &	10.99 \\
$\nu_{139}$ &	700.86 &	13.34 \\
$\nu_{141}$ &	669.61 &	5.86 \\
$\nu_{142}$ &	666.93 &	4.95 \\
$\nu_{143}$ &	662.60 &	3.74 \\
$\nu_{145}$ &	644.47 &	0.99 \\
$\nu_{146}$ &	641.23 &	3.91 \\
$\nu_{147}$ &	638.53 &	3.71 \\
$\nu_{150}$ &	614.37 &	0.68 \\
$\nu_{153}$ &	604.46 &	0.92 \\
$\nu_{155}$ &	595.78 &	0.51 \\
$\nu_{158}$ &	571.02 &	10.38 \\
$\nu_{160}$ &	563.04 &	8.70 \\
$\nu_{164}$ &	547.61 &	16.40 \\
$\nu_{166}$ &	525.86 &	2.78 \\
\hline \\
\end{tabular}
\end{minipage}
\begin{minipage}{0.45\textwidth}
\centering
\captionof{table}{{mode number, position and intensity of the vibrational modes in the 1000$-$500 cm$^{-1}$ range in singly-deuterated circumovalene. The location of the substituting D is at duet C-H site.}}
\label{tab:tab32}
\begin{tabular}[c]{c|c|c} \hline
mode & frequency & IR intensity \\
 & cm$^{-1}$ & km/mol \\
\hline 
$\nu_{125}$ &	907.26 &	20.33 \\
$\nu_{126}$ &	905.76 &	4.90 \\
$\nu_{127}$ &	903.02 &	1.73 \\
$\nu_{129}$ &	884.73 &	143.14 \\
$\nu_{130}$ &	884.48 &	0.56 \\
$\nu_{131}$ &	883.31 &	12.12 \\
$\nu_{133}$ &	874.35 &	17.55 \\
$\nu_{134}$ &	868.57 &	0.72 \\
$\nu_{135}$ &	860.35 &	1.57 \\
$\nu_{137}$ &	846.70 &	7.32 \\
$\nu_{143}$ &	813.43 &	0.91 \\
$\nu_{146}$ &	787.58 &	2.00 \\
$\nu_{147}$ &	780.75 &	27.70 \\
$\nu_{149}$ &	776.59 &	5.98 \\
$\nu_{151}$ &	772.26 &	3.49 \\
$\nu_{156}$ &	742.10 &	0.63 \\
$\nu_{158}$ &	738.09 &	6.26 \\
$\nu_{159}$ &	734.86 &	0.57 \\
$\nu_{162}$ &	721.92 &	0.61 \\
$\nu_{163}$ &	714.27 &	0.87 \\
$\nu_{165}$ &	706.84 &	5.87 \\
$\nu_{168}$ &	679.15 &	1.18 \\
$\nu_{169}$ &	677.95 &	1.56 \\
$\nu_{170}$ &	673.61 &	1.23 \\
$\nu_{171}$ &	671.02 &	3.32 \\
$\nu_{173}$ &	664.29 &	2.27 \\
$\nu_{176}$ &	632.09 &	1.07 \\
$\nu_{178}$ &	623.55 &	3.47 \\
$\nu_{180}$ &	609.46 &	3.86 \\
$\nu_{184}$ &	597.18 &	19.91 \\
$\nu_{185}$ &	595.46 &	4.09 \\
$\nu_{193}$ &	562.39 &	11.46 \\
$\nu_{198}$ &	540.46 &	1.05 \\
$\nu_{201}$ &	510.83 &	1.67 \\
$\nu_{202}$ &	508.63 &	1.74 \\
\hline \\
\end{tabular}
\end{minipage}
\begin{minipage}{0.45\textwidth}
\centering
\captionof{table}{{mode number, position and intensity of the vibrational modes in the 1000$-$500 cm$^{-1}$ range in doubly-deuterated circumovalene. The location of the substituting Ds are at duet C-H site.}}
\label{tab:tab33}
\begin{tabular}[c]{c|c|c} \hline
mode & frequency & IR intensity \\
 & cm$^{-1}$ & km/mol \\
\hline 
$\nu_{117}$ &	977.16 &	1.28 \\
$\nu_{123}$ &	948.21 &	1.90 \\
$\nu_{125}$ &	905.09 &	4.69 \\
$\nu_{127}$ &	885.32 &	1.26 \\
$\nu_{128}$ &	884.80 &	144.56 \\
$\nu_{129}$ &	883.38 &	14.88 \\
$\nu_{130}$ &	878.43 &	6.88 \\
$\nu_{131}$ &	876.42 &	22.15 \\
$\nu_{135}$ &	847.22 &	5.12 \\
$\nu_{138}$ &	840.80 &	0.88 \\
$\nu_{139}$ &	832.45 &	4.52 \\
$\nu_{145}$ &	787.60 &	2.54 \\
$\nu_{146}$ &	782.76 &	20.14 \\
$\nu_{148}$ &	776.56 &	7.30 \\
$\nu_{149}$ &	774.20 &	0.75 \\
$\nu_{150}$ &	773.24 &	4.51 \\
$\nu_{151}$ &	771.32 &	3.25 \\
$\nu_{157}$ &	738.56 &	4.90 \\
$\nu_{158}$ &	735.31 &	1.20 \\
$\nu_{163}$ &	710.56 &	0.95 \\
$\nu_{165}$ &	706.53 &	5.39 \\
$\nu_{167}$ &	690.64 &	0.77 \\
$\nu_{168}$ &	679.12 &	1.17 \\
$\nu_{169}$ &	676.06 &	4.29 \\
$\nu_{172}$ &	668.55 &	1.74 \\
$\nu_{173}$ &	660.99 &    2.20 \\
$\nu_{176}$ &	632.03 &	1.07 \\
$\nu_{178}$ &	622.59 &	3.25 \\
$\nu_{180}$ &	608.77 &	3.94 \\
$\nu_{182}$ &	603.52 &	1.57 \\
$\nu_{184}$ &	597.04 &	13.24 \\
$\nu_{185}$ &	592.53 &	5.55 \\
$\nu_{186}$ &	588.55 &	0.99 \\
$\nu_{188}$ &	576.09 &	0.97 \\
$\nu_{191}$ &	564.24 &	5.64 \\
$\nu_{194}$ &	558.16 &	5.78 \\
$\nu_{195}$ &	550.43 &	1.25 \\
$\nu_{197}$ &	540.59 &	1.17 \\
$\nu_{198}$ &	534.04 &	2.52 \\
$\nu_{200}$ &	512.29 &	0.62 \\
$\nu_{201}$ &	509.85 &	1.19 \\
$\nu_{202}$ &	507.82 &	1.12 \\
\hline \\
\end{tabular}
\end{minipage}
\bsp	
\label{lastpage}
\end{document}